\documentclass[fleqn,10pt,usenames,dvipsnames]{wlscirep}
\usepackage[utf8]{inputenc}
\usepackage{graphicx}
\usepackage{floatrow}
\usepackage[T1]{fontenc}
\usepackage[normalem]{ulem}
\usepackage{amssymb,amsfonts,amsmath}
\usepackage{xr-hyper} 
\usepackage{hyperref}
\usepackage{comment}
\usepackage{chngpage}
\usepackage{dcolumn}
\usepackage{booktabs,tabularx,ltablex,longtable}
\usepackage{eucal} 
\usepackage{paralist}
\usepackage{enumitem}
\usepackage{filecontents} 

\definecolor{MyBrick}{rgb}{0.84,0.01,0.01}

\newcommand{\edited}[1]{\textcolor{blue}}

\newcommand{\avg}[1]{\langle #1 \rangle}

\newcommand{\etal}{\emph{et al.}}
\newcommand{\ie}{\emph{i.e.,} }
\newcommand{\eg}{\emph{e.g.,} }

\newcommand{\kshell}{$k$-shell}
\newcommand{\kcore}{$k$-core}
\newcommand{\ks}{k_\text{s}}
\newcommand{\nc}{n_\text{C}} 
\newcommand{\NC}{N_\text{c}} 
\newcommand{\degswap}{\texttt{deg}}
\newcommand{\commaswap}{\texttt{commA}}
\newcommand{\commbswap}{\texttt{commB}}
\newcommand{\Louvain}{\texttt{Lvn}}
\newcommand{\LouvainR}{\texttt{LvnR}}
\newcommand{\SBM}{\texttt{SBM}}

\newcommand{\brite}{\textsc{brite}}

\graphicspath{{figures/}}

\title{Interplay between \kcore{} and community structure in complex networks}

\author[1]{Irene Malvestio}
\author[2,1,3,*]{Alessio Cardillo}
\author[4,5,1,*]{Naoki Masuda}

\affil[1]{Department of Engineering Mathematics, University of Bristol, Bristol, BS8 1UB, United Kingdom}
\affil[2]{Department of Computer Science and Mathematics, University Rovira i Virgili, E-43007 Tarragona, Spain}
\affil[3]{GOTHAM Lab -- Institute for Biocomputation and Physics of Complex Systems (BIFI), University of Zaragoza, E-50018 Zaragoza, Spain}
\affil[4]{Department of Mathematics, University at Buffalo, Buffalo, NY,14260-2900, United States}
\affil[5]{Computational and Data-Enabled Science and Engineering Program, University at Buffalo, State University of New York, Buffalo, NY 14260-5030, USA}

\affil[*]{alessio.cardillo@urv.cat; naokimas@buffalo.edu}

\begin{abstract}
The organisation of a network in a maximal set of nodes having at least $k$ neighbours within the set, known as \kcore{} decomposition, has been used for studying various phenomena. It has been shown that nodes in the innermost \kshell{s} play a crucial role in contagion processes, emergence of consensus, and resilience of the system. It is known that the \kcore{} decomposition of many empirical networks cannot be explained by the degree of each node alone, or equivalently, random graph models that preserve the degree of each node (\ie{} configuration model). Here we study the \kcore{} decomposition of some empirical networks as well as that of some randomised counterparts, and examine the extent to which the \kshell{} structure of the networks can be accounted for by the community structure. We find that preserving the community structure in the randomisation process is crucial for generating networks whose \kcore{} decomposition is close to the empirical one. We also highlight the existence, in some networks, of a concentration of the nodes in the innermost \kshell{s} into a small number of communities.
\end{abstract}

\begin{document}


\flushbottom
\maketitle

\thispagestyle{empty}

%
%

\section*{Introduction}

Whenever a system can be abstracted as a set of units (\emph{nodes}) interacting in pairs (\emph{edges}), we can describe it as a network (also called a graph). Network analysis has proven to be a valuable framework to aid us to understand a plethora of phenomena taking place in many complex systems. Examples include cascades and collective behaviour in socio-technical systems, the emergence of cognitive functions in neural systems, the stability of chemical/biological systems, and the shape of spatially embedded systems, to cite a few~\cite{boccaletti-phys_rep-2006,barabasi-nphys-2011,vespignani-nat_phys-2012}.

One of the advantages of the network representation is the possibility to probe the system in a coarse-grained manner, going beyond dyadic interactions by identifying high-order structures of the network~\cite{lambiotte-nat_phys-2019,benson-science-2016}. Examples include tightly connected groups of nodes, \ie{} communities~\cite{fortunato2016community}, multiscale coarse-grained structures~\cite{gfeller-prl-2007}, core-periphery structure~\cite{borgatti-soc_net-2000,csermely-jcomnet-2013,rombach-siam_rev-2017}, nested assembly of nodes~\cite{mariani-phys_rep-2019}, rich clubs~\cite{zhou-ieee-2004,colizza-nphys-2006}, and the \kcore{}~\cite{erdos-acta_math_hun-1966,seidman1983network}.

The \kcore{} decomposition of a network is the maximal set of nodes that have at least $k$ neighbours within the set~\cite{erdos-acta_math_hun-1966,seidman1983network}. The algorithm to extract the \kcore{} consists in recursively removing the nodes having less than $k$ connections. A \kshell{} is defined as the set of nodes belonging to the $k$\textsuperscript{th} core but not to the $(k + 1)$\textsuperscript{th} core~\cite{seidman1983network}. The \kcore{} decomposition has proven to be useful in a variety of domains such as identifying and ranking the most influential spreaders in networks, identifying keywords used for classifying documents, and in assessing the robustness of mutualistic ecosystem and protein networks~\cite{kong2019k,malliaros-vldb_jour-2020}.

Models to generate random networks with specific features should help us to understand how the mechanisms governing the establishment of edges account for properties of empirical networks. Despite the vast range of applications of the \kcore{} decomposition, to the best of our knowledge, there have been only few attempts to build models to generate networks with a given \kcore{} structure. One indirect attempt to generate networks with a given \kcore{} decomposition is the so-called \brite{} model~\cite{medina-mascots-2001}. Originally, this model sought to replicate the features (including the \kcore{}) of the Internet network at the Autonomous System (AS) level by mixing the mechanism of growth with preferential attachment~\cite{de_solla_price-science-1965,barabasi-science-1999} and that of adding edges between already existing nodes. Another model aimed at generating networks with a \kcore{} structure akin to an empirical one by leveraging the information stored in the so-called core fingerprint~\cite{baur-net_het_media-2008}. The core fingerprint corresponds to knowing the number of nodes in each \kshell{}, the number of intra-shell edges (\ie{} those connecting nodes belonging to the same \kshell{}), and the number of inter-shell edges (\ie{} those connecting nodes belonging to different \kshell{s}) of a given network. Moreover, the authors qualitatively compared the Internet AS networks and synthetic networks preserving the core fingerprint of the original networks using several indicators~\cite{baur-net_het_media-2008}. More recently, models based on modified versions of the so-called configuration model have been proven to be effective in generating networks with \kcore{} structure akin to that of empirical networks \cite{hebert_dufresne-pre-2013,allard-prx-2019}. In a nutshell, in these models the edge stubs attached to each node are divided into two groups: red and blue. Red stubs can create any edges regardless of the \kshell{} structure.  Blue stubs only form edges connecting nodes belonging to distinct \kshell{s}. Among the possible pairs of stubs' colours, only the blue-blue pair is forbidden.

As mentioned above, another type of mesoscale structure is communities. Although there is not a univocal definition of what a community is, in general the community refers to a group of nodes that are more tightly connected between each other than with the other nodes of the network~\cite{fortunato2016community}. Communities are also defined by the concept of stochastic equivalence, \ie{} nodes in the same group/community interact, on average, with nodes in other groups in the same way~\cite{young-pre-2018}. Methods based on different definitions of communities may return different partitions of the node set. However, there is often some consistency between those partitions, which indicates the presence of groups of nodes acting like the building blocks of communities~\cite{riolo-pre-2020}. The presence of communities is an important large-scale characteristic of many empirical networks because a system's different functions tend to be located in different communities (\eg{} in functional brain networks~\cite{meunier-front_neuro-2010} and protein-protein interaction networks~\cite{huttlin-nature-2017}). Moreover, it has been proven that communities play a role in the resilience of the system~\cite{gilarranz-science-2017} and the presence of triangles~\cite{orman-compl_nets-2013}, as well as in the emergence of collective behaviour including synchronisation~\cite{lotfi-chaos-2018}, the emergence of cooperation~\cite{fotouhi-j_rsoc_int-2019,giatsidis2011evaluating}, spreading of a pandemic~\cite{salathe-pcompbio-2010}, and the attainment of consensus~\cite{mistry-pre-2015,masuda-pre-2014}.

Although \kcore{} and communities are two ways of decomposing the same network, there may be overlaps or intricate relationships between them. In the present paper, we study the relation between the \kcore{} decomposition and the community structures of several empirical and synthetic networks. In particular, we leverage the work of Alvarez-Hamelin \etal{}~\cite{alvarez2008k} and confirm that the nodes' degrees (\ie{} their number of edges) alone are not capable of reproducing the network's \kshell{} structure. We find that one has to include information about the community structure to obtain networks whose \kcore{} decomposition looks sufficiently close to the empirical one. We also highlight the existence of a concentration-like phenomenon of the innermost \kshell{s} into a small number of communities, which is stronger in some data sets than others.

%
%

\section*{Results}

\subsection*{Degree-based reconstruction of the \kcore{}}

As stated above, various studies on networks leverage the \kcore{} decomposition to extract insightful information from networks. However, less studies have asked which mechanisms are sufficient for explaining generation of networks having empirically observed patterns of \kcore{} decomposition. More specifically, Alvarez-Hamelin \etal{} found that networks generated using the configuration model~\cite{fosdick-siam_rev-2018} having a Poisson or power-law distribution do not display a \kcore{} structure similar to the one displayed by the AS network~\cite{alvarez2008k}. Using the results of Alvarez-Hamelin \etal{} as a starting point, given an empirical network $G$ with $N$ nodes, we check whether its \kcore{} decomposition can be reproduced solely from the degree of each node $i$ (\ie{} the number of edges that node $i$ has), denoted by $k_i$. We generated random networks by a standard configuration model preserving the degree of each node of $G$, which we denote by \degswap{} (see Methods for details).

We have analysed several empirical networks encompassing social, technological, linguistic, and transportation systems whose main properties are summarised in Table~\ref{tab:data}. In Fig.~\ref{fig:cumR_all}, we show the survival function of the probability distributions of the \kshell{} index, $P_{\geq}(\ks)$ (\ie{} fraction of nodes whose \kshell{} index is larger than or equal to $\ks$), for a selection of data sets, compared across the original networks and their synthetic counterparts (see Supplementary Fig.~\ref{S-fig:cumR_other_data} in SM for the other data sets). Figure~\ref{fig:cumR_all} indicates that the degree of each node is not sufficient for reproducing the \kcore{} structure of the original networks because $P_{\geq}(\ks)$ for \degswap{} considerably deviates from that for the original networks. This result is consistent with the previous results~\cite{alvarez2008k}. In fact, we find that fixing the degree of each node is sufficient to recover the \kcore{} profile in some networks. For these networks the empirical and \degswap{} networks are not too different in terms of $P_{\geq}(\ks)$ (\eg{} Facebook 2 and Cookpad networks). We point out two main differences in $P_{\geq}(\ks)$ between the empirical and \degswap{} networks. First, for most data sets, the largest $\ks$ value, which is denoted by $D$ and called the degeneracy, is considerably smaller for the networks generated by \degswap{} than the original networks. Second, the $P_{\geq}(\ks)$ of some empirical networks have plateaus and abrupt drops in $\ks \le D$. The plateaus imply that some of the \kshell{s} are completely or almost empty, whereas the abrupt drops indicate that some \kshell{s} are more densely populated than those adjacent to them. In contrast, $P_{\geq}(\ks)$ for the \degswap{} networks does not have a notable plateau or drop in $\ks \le D$. Therefore, in the \degswap{} networks, all the \kshell{s} up to $\ks = D$ are populated, and there is no \kshell{} that is substantially more populated than its adjacent \kshell{s}.

A more quantitative comparison of distribution $P_{\geq}(\ks)$ between the empirical and \degswap{} networks may be done by, for example, the Kolmogorov-Smirnov (KS) test~\cite{massey1951kolmogorov}. However, because a majority of the nodes usually belongs to outer \kshell{s}, (\ie{} set of nodes with small $\ks$ values) and Fig.~\ref{fig:cumR_all} shows that the strongest discrepancies between the two distributions tend to occur at large $\ks$ values, the KS test fails to grasp the differences at large $\ks$ values that we are mostly interested in. Therefore, we compare the \kcore{} decomposition of the empirical and \degswap{} networks using four indicators, \ie{} the relative difference in the average \kshell{} index, $\Delta \avg{\ks}$, the relative difference in the network's degeneracy, $\Delta D$, the Jaccard score, $J$, and Kendall's, $\tau_K$ of the nodes belonging to the top 10\% (\ie{} innermost \kshell{s}) of the $P_{\geq}(\ks)$ distribution. The average of each indicator over all the data sets for the networks obtained with the \degswap{} shuffling method is equal to $\avg{\Delta\avg{\ks}} = 0.052 \pm 0.056$, $\avg{\Delta D} = 0.302 \pm 0.288$, $\avg{J} = 0.563 \pm 0.194$, and $\avg{\tau_K} = 0.763 \pm 0.176$. The value of $\avg{\Delta\avg{\ks}}$ indicates that $\avg{\ks}$ is only $\approx 5\%$ different between the original and \degswap{} networks on average. However, their degeneracy differs by $\approx 30\%$ on average. The $\avg{J}$ and $\avg{\tau_K}$ values inform us that innermost \kshell{s} of the original networks and those of the \degswap{} networks tend to share approximately half of the nodes, albeit their ranking seems to be fairly preserved. The values of each indicator are reported in Supplementary Table~\ref{S-tab:summary1}.

\subsection*{Community-aware reconstruction of the \kcore{}}

We have seen that the degree distribution by itself does not reproduce main features of the \kshell{} index distribution. An alternative feature that may explain the \kshell{} index distribution is the community structure. For this reason, we generated synthetic networks that preserve both the degree of each node and the community structure, $\mathcal{C} = \{ C_1, \ldots, C_{\NC} \}$, where $\NC$ is the number of communities of the original network. To account for the multiple definitions of what a community is, we identified the communities of each network using two methods: the Louvain method~\cite{blondel-jstat-2008}, denoted by \Louvain{}, and the degree-corrected stochastic block model~\cite{karrer2011stochastic}, denoted by \SBM{}. In combination with each of the two community detection methods, we considered two rewiring methods preserving $\mathcal{C}$ and the degree of each node, denoted by \commaswap{} and \commbswap{}. Method \commaswap{} preserves the exact number of inter- and intra-community edges at the level of single communities. Method \commbswap{} preserves the number of inter- and intra-community edges for each node.

Figure~\ref{fig:cumR_all} indicates that preserving the community structure in addition to the degree of each node improves the similarity in $P_{\geq}(\ks)$ between the empirical and synthetic networks, especially at large $\ks$ values, which correspond to inner \kshell{s}. In particular, \commaswap{} and \commbswap{} generate networks whose $D$ value tends to be closer to the empirical value than \degswap{} does. Furthermore, $P_{\geq}(\ks)$ for \commaswap{} and \commbswap{} tends to have plateaus and abrupt drops at $\ks \le D$ similarly to the empirical networks. Overall, synthetic networks preserving the \SBM{} community structure have a \kcore{} decomposition more akin to the empirical one than those preserving the \Louvain{} community structure. This observation is quantitatively supported by the values of the four indices reported in Supplementary Table~\ref{S-tab:summary1}. 

\begin{figure}[ht]
\centering
\includegraphics[width=.8\linewidth]{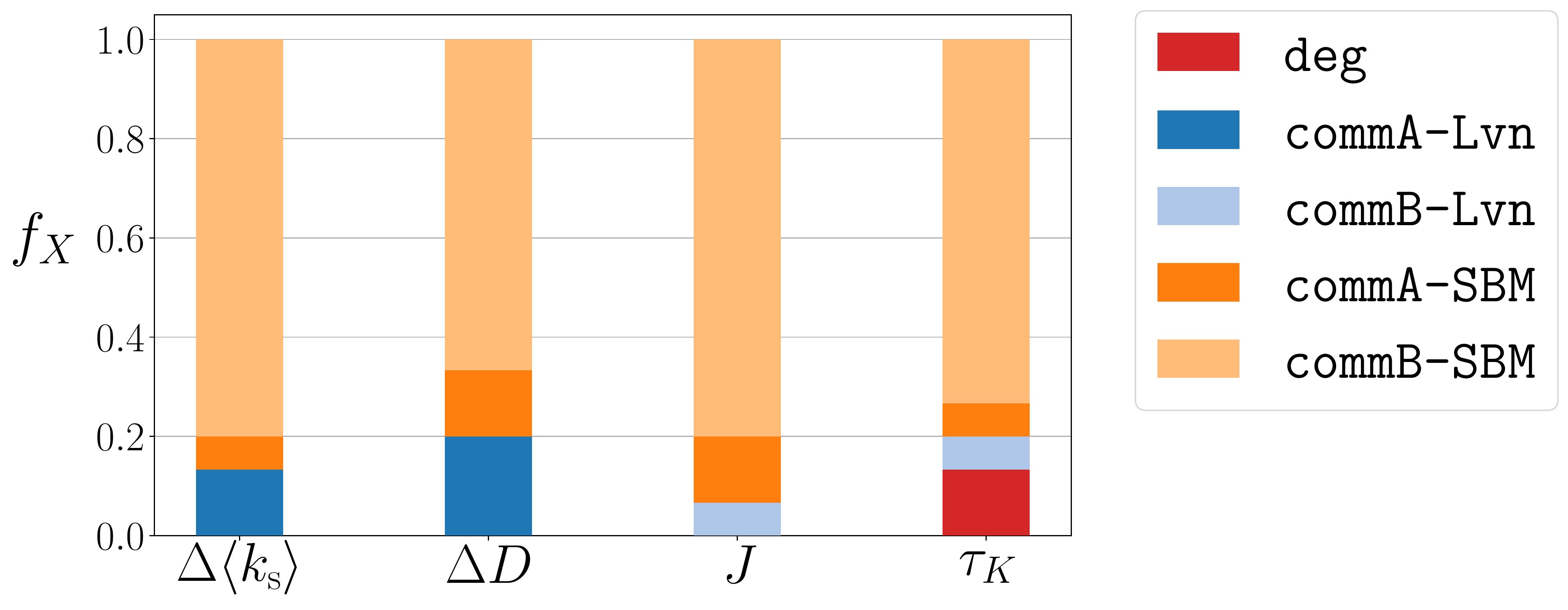}
\caption{Performances of different shuffling methods in terms of four indicators. We report the fraction of data sets for which a given combination of the shuffling method and the community detection method yields an indicator's value closest to that for the original network. Each bar refers to an indicator, \ie{} average \kshell{'s} difference, $\Delta\avg{\ks}$, degeneracy's difference, $\Delta D$, Jaccard score, $J$, and Kendall's tau, $\tau_K$.}
\label{fig:summary_performance}
\end{figure}

To obtain an overview of the performances of different network randomisation methods, in Fig.~\ref{fig:summary_performance} we show the fraction of data sets, $f_X$, for which a certain shuffling method generates a \kcore{} decomposition that is the most similar to that of the empirical network according to each indicator. The figure indicates that \commbswap{-\SBM{}} (\ie the \commbswap{} shuffling method that preserves the community structure determined by \SBM{}) performs the best in mimicking the \kshell{} index features for approximately 65\%--80\% of the data sets, depending on the indicator. Detailed results for the performance of each method for each empirical network are shown in Supplementary Fig.~\ref{S-fig:summary_difference} and Supplementary Table~\ref{S-tab:summary1}.

One issue of \Louvain{} is that it cannot discover small communities~\cite{fortunato2007resolution,fortunato2016community}. One way to mitigate this limitation is to introduce in \Louvain{} a resolution parameter, $r \in (0,1]$, regulating the resolution scale. It is possible to detect small communities when $r$ is small, whereas the original \Louvain{} corresponds to $r=1$~\cite{lambiotte-ieee_trans_netsci_eng-2014}. We denote by \LouvainR{} the Louvain method with $r<1$, \ie{} with a resolution higher than that used by \Louvain{}. In Sec.~\ref{S-sec:louvain_resolution} of SM we report whether preserving the communities found using \LouvainR{} instead of \Louvain{} improves our ability to reproduce the \kcore{} decomposition of the original network. We found that \LouvainR{} performs better than \Louvain{} (Supplementary Figs.~\ref{S-fig:cumR_Lr3} and \ref{S-fig:summary_difference_Lr}) but worse than \SBM{} in general (Supplementary Fig.~\ref{S-fig:summary_performance_Lr}).

Imposing the simultaneous conservation of each node's degree and community structure may result in synthetic networks that are not substantially different from the original ones. To exclude this possibility, we computed the Jaccard score, $J(\mathcal{L},\mathcal{L}^\prime)$, (see Eq.~\eqref{eq:jaccard}) for the sets of edges, $\mathcal{L}$ and $\mathcal{L}^\prime$, of the original and shuffled networks, respectively. The values of $J$ approximately fall between $0.01$ and $0.5$, confirming that the set of edges -- hence, the networks -- are considerably different.

The results presented so far suggest that preserving the community structure improves the preservation of the \kcore{} decomposition of the original network. Therefore, the mere presence of a community structure may be enough to preserve the main features of the \kcore{} decomposition of the original networks. To test this possibility, we applied the \kcore{} decomposition to networks with communities generated using the LFR model~\cite{lancichinetti-pre-2008} (see Sec.~\ref{S-sec:lfr} of SM). The plots of $P_{\geq}(\ks)$ shown in Supplementary Figs.~\ref{S-fig:cumR_LFR_G1}--\ref{S-fig:cumR_LFR_G4} indicate that the presence of a community structure alone is not sufficient for producing major features of the \kcore{} structure of the empirical networks. Specifically, the $P_{\geq}(\ks)$ of the networks generated by the LFR model is always smooth and shows neither plateaus nor abrupt drops as $\ks$ increases. Moreover, with the LFR, $\ks$ is narrowly distributed, \ie{} $\max(\ks) - \min(\ks) \approx 10$. These differences between the \kcore{} structure of the LFR model and that of empirical networks  are not sensitive to the value of the mixing parameter, $\mu$, of the LFR model, which controls how distinct the communities are. It should also be noted that for the LFR model, as for the empirical network, the \commbswap{-\SBM{}} generates networks that are the most similar to the original LFR networks among the different shuffling methods in terms of $P_{\geq}(\ks)$.

\subsection*{Overlap between communities and \kcore{}}

Preserving the community structure in addition to the node's degree can lead to preservation of features of the \kcore{} structure possibly because nodes with high values of $\ks$ form a \kcore{} which tend to belong to the same community. To examine this possibility, we show the number of communities to which the nodes of a given \kshell{} belong, $\nc(\ks)$, in Fig.~\ref{fig:nrcomms_vs_kshell} (see Supplementary  Fig.~\ref{S-fig:nrcomms_vs_kshell_alldata} for the other data sets). Although each data set shows a distinct pattern, for many data sets, inner \kshell{s} (\ie{} nodes with large $\ks$ values) are concentrated into one or a few communities. The concentration effect is particularly noticeable for some data sets, \eg{} Facebook 1 and Twitter. To check whether the number of communities per \kshell{} is merely a byproduct of the random combinatorial effect owing to the number of communities, the distribution of the community size, and the distribution of $\ks$, we computed a random assignments of the nodes to communities and then calculated $\nc(\ks)$ for each $\ks$ value (see Sec.~\ref{S-sec:rel_comms_kcore} and Supplementary Fig.~\ref{S-fig:ncomm_kshell_richclub_purecomb} of the SM). We have found that the nodes in each \kshell{} are almost always more concentrated into a smaller number of communities than what is expected by the random assignment of the nodes to communities for all the data sets and community detection methods, with the only exception of \SBM{} for Cookpad's data sets. This finding is in agreement with the previous result that nodes with high $\ks$ tend to belong to the same community, which has been observed in networks embedded into hyperbolic spaces~\cite{faqeeh-prl-2018,osat-prr-2020}. In particular, we observe a strong concentration of the \kshell{s} into a few communities for the Facebook 1, Twitter, Cond. Matter, Comp. Science, and Words networks, which are those showing a more pronounced difference in the values of $D$ between the original and \degswap{} networks.

%
%

\section*{Discussion}

The information encoded in the degree of each node is not sufficient for generating networks with a \kcore{} structure that is similar to those of empirical networks~\cite{alvarez2008k}. This gap of knowledge calls for the design of generative models of networks beyond the configuration model. Such models are expected to be useful to generate benchmark networks and to understand the mechanisms behind the emergence of the \kcore{}. To the best of our knowledge, few models are available to generate networks with a given \kcore{} decomposition~\cite{baur-net_het_media-2008,hebert_dufresne-pre-2013,allard-prx-2019}.

In the present study, we investigated how much the combination of the nodes' degrees and community structure accounts for \kcore{} structure of empirical networks. Given a network $G$, we randomly shuffled $G$'s edges to generate its synthetic counterparts preserving each node's degree and/or community structure of $G$. We found that randomised networks preserving the community structure obtained through a stochastic block model showed a \kshell{} index distribution that was reasonably similar to the distribution for the original networks. The success of the stochastic block model in mimicking the features of the \kcore{} decomposition might be due to its ability to approximate the mesoscale structures of networks with a good accuracy~\cite{olhede-pnas-2014,young-pre-2018}, including communities. We also sought to understand more the relationship between \kcore{} and communities by studying networks generated by the LFR model which enables us to control the extent to which the communities are distinguished from each other. However, regardless of whether or not different communities are relatively distinguished from each other in a network, the \kshell{} index distribution of LFR networks does not show the same features as those observed in the empirical networks. Finally, we have investigated the overlap between communities and \kshell{s} and found that, in some empirical networks, the nodes in inner \kshell{s} are concentrated into a small number of communities much more so than a randomised counterpart. This result is in agreement with the observations made for networks embedded in hyperbolic spaces~\cite{faqeeh-prl-2018,osat-prr-2020}. Up to our numerical efforts, the concentration is observed if and only if the empirical network and its \degswap{} counterpart are substantially different in terms of their \kcore{} decomposition. The concentration suggests that inner \kshell{s} may perform specific functions in such networks, corresponding to the functions of the communities they belong to as observed in, for instance, functional brain networks~\cite{meunier-front_neuro-2010} and protein-protein interaction networks~\cite{huttlin-nature-2017}.

The ``community aware'' rewiring mechanisms introduced in this paper can be used for assessing whether or not a given property of a network is a direct expression of its community structure. One example of such an approach is given in~\cite{mozafari2019improving}, where the authors have improved the robustness against attacks on a network while keeping its community structure. In that case, the method only preserves the communities and alters the connectivity pattern by increasing the density of intra-community edges as well as changing the edges between communities. It may be interesting, instead, to check whether the robustness of the network can be improved even when one also preserves the degree of the nodes using our community-aware rewiring mechanisms.

One viable extension of our work is to the case of $k$-peak graph decomposition method~\cite{govindan2017k}. In Ref.~\cite{govindan2017k}, the authors argue that for networks with communities, the \kcore{} decomposition should be performed locally rather than globally, thus returning the $k$-peak decomposition of each of the system's regions. The rationale behind this approach is to avoid that, if the network contains regions with different densities of edges, the standard \kcore{} decomposition would fail to recognise local core nodes in sparser regions. Studying the evolution of the $k$-peak decomposition in response to the rewiring of the connections may unveil salient features of complex systems. Another possible direction of research is to concatenate the information encoded into the \kshell{} index, $\ks$, with the one provided by the so-called onion decomposition (OD)~\cite{hebert_dufresne-scirep-2016}. The OD is an extension of the \kcore{} decomposition where a node is labelled with both its $\ks$ and its layer index. The layer of a node $i$ represents the iteration number with which node $i$ is removed in the recursive pruning process of the \kcore{} decomposition. The OD provides a further characterisation of the structure of the network than the \kcore{}, revealing, for example, how tree-like the network is.

Summing up, in this work we have analysed the interplay between the \kcore{} decomposition and community structure of networks. Understanding such a relationship is useful not only owing to the broad range of applications of \kcore{} decomposition, but also to inform the design of models capable of generating networks with both a community structure and \kcore{'s} features beyond those explainable by the degree distribution. Such models may stand on, for instance, the stochastic block model~\cite{karrer2011stochastic}, the enhanced configuration model based on maximum entropy~\cite{mastrandrea-njp-2014}, or the hierarchical extension of the LFR model~\cite{yang-pre-2017}. Alternatively, models based on microscopic growth mechanisms such as triadic closure~\cite{kumpula-prl-2007,bianconi-pre-2014} or modified preferential attachment~\cite{shang-chaos-2020} may deserve further investigation.

%
%

\section*{Methods}

\subsection*{Data}

We have considered networks corresponding to systems of different types: from social to technological, from semantic to transportation. Table~\ref{tab:data} summarises main properties of such networks. Except for Cookpad networks, all the data sets are publicly available and have been retrieved from the Stanford Large Network data set Collection~\cite{leskovec2012learning} (Facebook 1, Twitter, Emails, and Cond. Matter), the Network Repository~\cite{nr, traud2012social, Traud:2011fs}  (Facebook 2, 3, 4, and 5), the Koblenz Network Collection (KONECT)~\cite{konect} (Comp. Science, and Words), Mark E. J. Newman's personal network data repository~\cite{pol_blog} (Web-blogs), and the OpenFlights data repository~\cite{airport_web} (Global airline). In the following text, we provide a brief description of each data set.
\begin{description}[leftmargin=0.5cm,labelindent=0.5cm]
    \item[Facebook \& Twitter.] These networks describe social relationships. Nodes are people. Edges represent their friendship relations.
    \item[Web-blogs.] This network is composed of the hyperlinks (edges) between weblogs on US politics (nodes) recorded in 2005.
    \item[Emails.] This is a network of email data from a large European research institution. Nodes are people. Edges connect pairs of individuals who have exchanged at least one e-mail.
    \item[Cond. Matter \& Comp. Science.] The former network is the co-authorship network of the authors of preprint manuscripts submitted to the Condensed Matter Physics  \texttt{arXiv} e-print archive from January 1993 to April 2003. The latter network is similarly defined using manuscripts appearing in the DBLP computer science bibliography, using a comprehensive list of research papers in computer science. The submission time of the papers of the DBLP collection is unavailable. A node is an author. An edge represents the existence of at least one manuscript co-authored by two authors.
    \item[Global airline.] In this network nodes are airports across the globe. An edge indicates direct commercial flights between two airports.
    \item[Words.] This network accounts for the lexical relationships among words extracted from the WordNet data set. Nodes are English words. Edges are relationships (synonymy, antonymy, meronymy, etc.) between pairs of words.
    \item[Cookpad.] These networks are extracted from the Cookpad online recipe sharing platform~\cite{cookpad-web}. Users can post and browse recipes, as well as interact with other users through recipes in multiple ways including liking, sharing, and posting a comment. The platform is present in many countries (\eg{} Japan, Indonesia, United Kingdom, and Italy). Here, we consider the data collected from September to November of 2018 in Greece, Spain, and the United Kingdom, separately for each country. In the three networks, nodes are users. An edge between a pair of users exists if one or more of the following types of events takes place: like or follow a user, viewing, bookmarking, commenting, or making a cooksnap of another user's recipe.
%
\end{description}
All the networks considered in this work are treated as undirected and unweighted, even when the original data contains more information. Finally, we also consider synthetic networks, generated using the LFR (Lancichinetti–Fortunato–Radicchi) model~\cite{lancichinetti-pre-2008}  (see Sec.~\ref{S-sec:lfr} of SM for details).

\subsection*{Network shuffling}

Given a network, $G$, with $N$ nodes and $L$ edges, we generate a randomised counterpart, $G^\prime$, that has the same nodes and the same number of edges by shuffling the edges of $G$. We consider three shuffling methods denoted by \degswap{}, \commaswap{}, and \commbswap{}; each shuffling method preserves different properties of $G$. The shuffling consists in selecting uniformly at random two edges $(a,b)$ and $(c,d)$, and replacing them with, \eg $(a,c)$ and $(b,d)$, if the swapping of the edges is accepted. An attempt to swap edges is accepted, in which case we call the swapping effective, if and only if it respects the rule of the specific shuffling method and the swapping does not generate self-loops or multiple edges. We continued the shuffling until we carried out $2L$ effective swaps, such that an edge was swapped four times on average. 

In the following text, we provide the details of each shuffling method. Assume that network $G$ partitions into communities such that the set of the communities is $\mathcal{C} = \{C_1, \ldots, C_{\NC} \}$, where $\NC$ is the number of communities. Furthermore, let\newline $g(i) \in \mathcal{C}, \; i = 1, \dots, N$, be the community to which the $i$th node belongs and $k_i$ be the degree of node $i$. We have:
\begin{description}
  \item[Degree-preserving shuffling (\degswap{}).] This method preserves degree $k_i$ of each node $i$ and is equivalent to the configuration model~\cite{fosdick-siam_rev-2018}. 
  \item[Community-preserving shuffling of type A (\commaswap{}).] On top of the degree of each node, this method preserves the total number of edges within each community and between each pair of communities. In attempts to swap edges, we replace two randomly selected edges $(a,b)$ and $(c,d)$ by $(a,c)$ and $(b,d)$ if and only if an end node of edge $(a,b)$ and an end node of edge $(c,d)$ belong to the same community (\ie if $g(b) = g(c)$ or $g(a) = g(d)$).
  \item[Community-preserving shuffling of type B (\commbswap{}).] Like \commaswap{}, this method preserves the degree of each node and the number of edges within each community and between each pair of communities. In contrast with \commaswap{}, the \commbswap{} method preserves the numbers of edges within and across communities for each node, and not only for each community or pairs of communities. Given two selected edges $(a,b)$ and $(c,d)$, we replace them with $(a,c)$ and $(b,d)$ if and only if the two new edges connect the same community pairs as before the swapping (\ie $g(b) = g(c)$ and $g(a) = g(d)$).
\end{description}

\subsection*{Comparison of the \kcore{} decomposition}

To assess the similarity between the \kcore{} decomposition of the original network, $G$, and of its shuffled counterpart, $G^\prime$, we used four indicators: the average \kshell{} index, $\avg{\ks}$, the network's degeneracy, $D$, the Jaccard score, $J$, and the generalised Kendall's tau, $\tau_K$. The indicator $\avg{\ks}$ explicitly depends on all the nodes in the network, whereas $D$, $J$ and $\tau_K$ only depend on the nodes belonging to the innermost \kshell{(s)}. We use the latter three indicators because, although a majority of nodes tends to belong to outer \kshell{s}, it is a difference in the tails of the $\ks$ distributions that often affect functions of networks such as the impact of influencers in contagion processes~\cite{kitsak-nphys-2010}. The four indicators are defined as follows.\newline
The average of the \kshell{} index, $\avg{\ks}$, is equal to
\begin{equation}
\label{eq:avg_ks}
\avg{\ks} = \frac{1}{N} \sum\limits_{i=1}^N \ks(i)\,,
\end{equation}
where $\ks(i)$ is the \kshell{} index of node $i$. 
The degeneracy, $D$, of a network $G$ is given by~\cite{bollobas1998modern}
\begin{equation}
\label{eq:degeneracy}
D = \max_{i \in G} \{ \ks(i) \} \,.
\end{equation}
Rather than using these raw indicators, to compare across the different data sets, we compute their relative difference between the empirical network and its shuffled counterpart given by $\Delta X = \left\lvert X_G - X_{G^\prime} \right\rvert/ X_G$, where $X \in \{  \avg{\ks} , D \}$. 

To compute $J$ and $\tau_K$, we need to define a criterion to select nodes belonging to the innermost \kshell{s}. We decided to confine the comparison to the nodes whose $\ks$ falls within the top $10\%$ among the $N$ nodes. The horizontal lines in Fig.~\ref{fig:cumR_all} indicate the threshold values of $\ks^\star$ such that $P_{\geq}(\ks^\star) = 0.1$. In the same manner, we define ${\ks^\star}^\prime$ such that $P_{\geq}({\ks^\star}^\prime) = 0.1$ in network $G^\prime$. To calculate $J$ and $\tau_K$, we use the nodes belonging to \kshell{s} with $\ks \ge \ks^\star$ in $G$ and the nodes belonging to \kshell{s} with $\ks \ge {\ks^\star}^\prime$ in $G^\prime$ without duplication of the nodes. There are several remarks. First, it may hold that $\ks^\star \neq {\ks^\star}^\prime$. Second, the value of ${\ks^\star}^\prime$ varies from one combination of a run of shuffling and community detection to another. Third, as in the case of the Facebook 2 data set, ${\ks^\star}^\prime$ sometimes does not even exist. In such a case, we set ${\ks^\star}^\prime = D$ and select all the nodes belonging to the innermost \kshell{} although they constitute more than 10\% of the nodes in the network. Fourth, additional tests using different threshold percentages, 5\% and 20\%, instead of 10\%, did not qualitatively change the results. Fifth, while the Jaccard score simply compares the nodes belonging to two sets, the generalised Kendall's tau, $\tau_K$ compares ranked sets. In our case, the node's rank is equivalent to the $\ks$ value.

Given two sets $\mathcal{A}$ and $\mathcal{B}$, the Jaccard score quantifies their overlap and is given by
\begin{equation}
\label{eq:jaccard}
J(\mathcal{A},\mathcal{B}) = \dfrac{\lvert \mathcal{A} \cap \mathcal{B}\rvert}{\lvert \mathcal{A} \cup \mathcal{B}\rvert}\,.
\end{equation}
The Jaccard index ranges between 0 and 1. A value of 1 indicates the complete overlap between the two sets (\ie the sets are the same), whereas a value of 0 indicates that the sets are completely different.

The generalised Kendall's tau, $\tau_K$, measures the consistency between two rankings by assigning penalties to pairs of elements on which the two rankings disagree~\cite{fagin2003comparing,mccown2007agreeing}. Given two sets $\mathcal{A}$ and $\mathcal{B}$ having $m_A$ and $m_B$ elements, respectively, consider their associated ranking functions $\mathcal{X}$ and $\mathcal{Y}$. We denote with $(z_1, z_2)$ an arbitrary pair of elements of $\mathcal{A} \cup \mathcal{B}$. We assign a penalty $K_{z_1,z_2}(\mathcal{X},\mathcal{Y}) = 1$ to $(z_1, z_2)$ if \begin{inparaenum}[(a)] \item the rankings of the two elements within each set are different (\ie $\mathcal{X}(z_1) \gtrless \mathcal{X}(z_2)$ and $\mathcal{Y}(z_1) \lessgtr \mathcal{Y}(z_2)$), \item the element with the higher rank in one set is missing in the other set, \ie $\mathcal{X}(z_1) > \mathcal{X}(z_2)$ and $z_1 \notin \mathcal{B}$ (or $\mathcal{X}(z_2) > \mathcal{X}(z_1)$ and $z_2 \notin \mathcal{B}$), or \item both elements belong to one set each, which is not the same set, \ie $z_1 \notin \mathcal{B}$ and $z_2 \notin \mathcal{A}$ (and vice-versa)\end{inparaenum}. In all the other cases $K_{z_1,z_2}(\mathcal{X},\mathcal{Y}) = 0 $, such that we do not penalise the $(z_1, z_2)$ pair. Finally, we sum the penalties over all the possible pairs of elements and normalise it, thus obtaining the generalised Kendall's tau:
\begin{equation}
\label{eq:gen_kendall}
\tau_K(\mathcal{X},\mathcal{Y}) = 1 - \frac{1}{m_A  m_B} \sum_{z_1, z_2 \in \mathcal{A} \cup \mathcal{B}} K_{z_1,z_2}(\mathcal{X},\mathcal{Y}).
\end{equation}
Index $\tau_K$ ranges between 0 and 1. If $\tau_K = 1$, the two rankings are completely coherent. If $\tau_K = 0$, the two sets $\mathcal{A}$ and $\mathcal{B}$ have no pair of elements on which rankings $\mathcal{X}$ and $\mathcal{Y}$ are coherent. The above formulation of the Kendall's tau is the so-called the optimistic approach~\cite{fagin2003comparing}. This means that we do not penalise the case in which a pairs of elements is present in one set and not in the other set.

\subsection*{Community detection methods}

We considered two methods for community detection. The first is the Louvain method (\Louvain{})~\cite{blondel-jstat-2008}, which is a heuristic greedy multiscale method that approximately maximises the modularity function. Given a network with $N$ nodes distributed among $\NC$ communities, the modularity, $Q$, reads
\begin{equation}
\label{eq:modularity}
Q = \dfrac{1}{2L} \sum_{i,j = 1}^N \left[ a_{i,j} - \dfrac{k_i k_j}{2L} \right] \delta\bigl(g(i),g(j)\bigr) \,,
\end{equation}
where $a_{i,j}$ is the element of the network's adjacency matrix $A$; $g(i)$ is the community to which the $i$-th node belongs ($1 \le g(i) \le \NC$), and $\delta\bigl(g(i),g(j)\bigr)$ is the Kronecker delta. A large value of $Q$ implies a good partitioning. The Louvain method seeks the partitioning that maximises the modularity. Note that we obtain $Q \approx 0$ for random assignment of nodes to communities and that we obtain $Q \approx 1$ when the network is made of perfectly disjoint communities.

The other community detection method that we used is the stochastic block model~\cite{holland-soc_net-1983}. It uses the probabilities $\mathcal{P} = \{ p_{C_i,C_j} \}$ with which there exists an edge $(a,b)$ connecting an arbitrarily selected node $a$ in community $C_i$ (\ie{} $g(a) = C_i$) and an arbitrarily selected node $b$ in community $C_j$ (\ie{} $g(b) = C_j$). Different instances of probabilities $\mathcal{P}$ allow the description of different mixing patterns. When the diagonal entries of $\mathcal{P}$ predominate, we obtain the most usual community structure, whereas other instances yield other structures such as bipartite or core-periphery structure.

To find the optimal partition, one maximises the likelihood function with respect to $\{ p_{C_i,C_j} \}$ corresponding to the partitioning $\mathcal{C} = \{ C_i \}$, where $i,j \in 1,\ldots,\NC$. The unnormalised log-likelihood, $\mathfrak{L}$, with which a partition of network $G$ into $\NC$ communities, $\mathcal{C}$, is reproduced reads
\begin{equation}
\label{eq:loglikelyhood}
\mathfrak{L} \bigl( G \, \bigl\vert \, \mathcal{C} \bigr. \bigr) = \sum_{i,j = 1}^{\NC} e_{ij} \, \log \left( \frac{e_{ij}}{m_i \, m_j} \right) \,,
\end{equation}
where $e_{ij}$ is the number of edges connecting community $C_i$ and community $C_j$, and $m_i$ is the number of nodes belonging to $C_i$.

The above formulation, however, has one major limitation: it assumes that the degrees of the nodes are distributed according to a Poisson-like function. To account for the degrees' heterogeneity, Karrer \etal{} have implemented the so-called degree corrected stochastic block model, in which the expected degree of each node is kept constant via the introduction of additional parameters~\cite{karrer2011stochastic}. Let $e_i$ be the sum of the node's degree over all nodes in community $C_i$. Then, the unnormalised log-likelihood for the degree-corrected stochastic block model reads
\begin{equation}
\label{eq:loglikelyhoodDC}
\mathfrak{L}_{\text{DC}} \bigl( G \, \bigl\vert \, \mathcal{C} \bigr. \bigr) = \sum_{i,j = 1}^{\NC} e_{ij} \, \log \left( \frac{e_{ij}}{e_i \, e_j} \right) \,.
\end{equation}

Equations~\eqref{eq:loglikelyhood} and \eqref{eq:loglikelyhoodDC} depend on the number of communities $\NC$. Because the value of $\NC$ is not known a priori, it is inferred through the minimisation of a quantity called the description length. The minimum description length principle describes how much a model compresses the data and allows us to find the optimal number of communities while avoiding overfitting~\cite{peixoto-pre-2017}. In the present work we use the degree-corrected stochastic block model and its implementation available in the Python Graph-tool package~\cite{peixoto_graph-tool_2014}, which we refer to as \SBM{} for brevity.


%
%

\bibliography{biblio}

%
%

\section*{Acknowledgements}

The authors thank A. Barrat for helpful comments on a preliminary version of this work. A.C. acknowledges the support of the Spanish Ministerio de Ciencia e Innovacion (MICINN) through Grant IJCI-2017-34300. I.M., A.C., and N.M. acknowledge the support of Cookpad Limited. This work was carried out using the computational facilities of the Advanced Computing Research Centre, University of Bristol -- \url{http://www.bristol.ac.uk/acrc/}. Numerical analysis has been carried out using the NumPy, NetworkX, and Graph-tool Python packages~\cite{oliphant-book-2006,vanderwalt-compscieng-2011,hagberg-scipy-2008,peixoto_graph-tool_2014}. Graphics have been prepared using the Matplotlib Python package~\cite{hunter-matplotlib-2007}.

\section*{Author contributions statement}

I.M. analysed data, designed the experiments, and performed simulations, A.C. analysed the results and designed the experiments. All authors discussed the methods and results, and wrote and reviewed the manuscript. 

\section*{Additional information}

\begin{description}
\item[Competing interests] The authors declare no competing interests.
\item[Data availability] The data sets on Cookpad\texttrademark analysed in the current study are not publicly available due to exclusive ownership of Cookpad Limited. All the other data sets are available from the corresponding repositories listed in the bibliography.
\end{description}

%
%

\newpage

%
%

\begin{table}[ht]
\centering
\newcolumntype{d}[1]{D{.}{.}{#1} }
\begin{tabular}{|l|ccd{3}cd{3}c|cccc|c|}
 \hline 
Data set & $N$ & $L$ & \multicolumn{1}{c}{$\langle k \rangle$} & $k_{\max}$ & \multicolumn{1}{c}{$\langle \ks \rangle$} & $D$ & $\NC^{\text{\Louvain{}}}$ & $Q^\text{\Louvain{}}$ & $\NC^{\text{\SBM{}}}$ & $Q^\text{\SBM{}}$  & Ref. \\\hline\hline
Facebook 1 & 4039 &  88234 & 43.691 & 1045 & 26.880 & 115 & 16 & 0.835   & 62 &  0.551  &\cite{face1, leskovec2012learning} \\
Facebook 2 & 6386 & 217662 & 68.168 & 930 & 35.712 & 56 & 19 & 0.419 & 198 &  0.158 & \cite{face2, nr,traud2012social,Traud:2011fs}\\
Facebook 3 & 2235 & 90954 & 81.391 &  467 & 44.508 & 63 & 8 & 0.436 & 87 & 0.139 &\cite{face3, nr,traud2012social,Traud:2011fs}\\
Facebook 4 & 11247 &  351358 & 62.480 &  415 & 32.413 & 63 & 10 & 0.438 & 274 & 0.193 & \cite{face4, nr,traud2012social,Traud:2011fs}\\
Facebook 5 &  27737 & 1034802 & 74.615 & 2555 &  38.681 & 81 & 18 & 0.470 & 547 & 0.172 & \cite{face5, nr,traud2012social,Traud:2011fs}\\
Twitter  &  81306 & 1342296 & 33.018 & 3383 &  17.762 & 96 & 73 & 0.808 & 510 & 0.511 & \cite{twitter, leskovec2012learning}\\
Web-blogs &  1490 & 16715 &   22.436  &  351 & 12.154 & 36 & 275 & 0.426 & 17 & 0.076 & \cite{pol_blog,adamic2005political}\\
Emails &  1005 & 16064 & 31.968 & 345 & 17.063 & 34 & 26 & 0.410 &  33 & 0.232 & \cite{email,yin2017local,leskovec2007graph}\\
Cond. Matter & 23133 &  93439 &  8.078  & 279 &  4.900 & 25 & 619 & 0.730 & 203 & 0.633 & \cite{condmat, leskovec2007graph}\\
Comp. Science & 317080 & 1049866  & 6.622 & 343 &  4.215  & 113 & 209 & 0.822 &  676 & 0.726 & \cite{konect, konect:2017:com-dblp, konect:leskovec2012}\\
Global airline & 3376 & 19179 &  11.362 & 248 &  6.123 & 31 & 26 & 0.665 & 40 & 0.311 &\cite{airport_web}\\
Words & 146005 & 656999 & 9.000 & 1008 & 5.289  & 31 & 378 & 0.759 & 548 & 0.583 & \cite{konect, konect:2017:wordnet-words,konect:fellbaum98}\\
Cookpad Greece &   32235 & 745178 &  46.234  &  8196 &  23.709 & 158 & 40 & 0.166 & 76 & 0.020 & -- \\
Cookpad Spain & 122158 & 1749751 & 28.647 & 12637  & 14.547  & 162 & 262 & 0.270 & 90 & 0.035 & --\\
Cookpad UK &  13758 & 47525 & 6.909 &  1880 &  3.558 & 33 & 199 & 0.350 & 8 & 0.114 & --\\
\hline 
\end{tabular}
\caption{Main properties of the data sets used in the present study. $N$: number of nodes, $L$: number of edges, $\avg{k}$: average degree, $k_{\max}$: maximum degree, $\avg{ks}$: average value of the \kshell{} index, $D$: maximum value of the \kshell{} index, $\NC^{\text{\Louvain{}}}$, $Q^\text{\Louvain{}}$: number of communities determined by the Louvain method and the corresponding modularity, respectively, $\NC^{\text{\SBM{}}}$, $Q^\text{\SBM{}}$: number of communities determined by the \SBM{} and the corresponding modularity, respectively.}
\label{tab:data} 
\end{table}
%

%
%
%
\begin{figure}[ht]
\centering
\includegraphics[width=1.\linewidth]{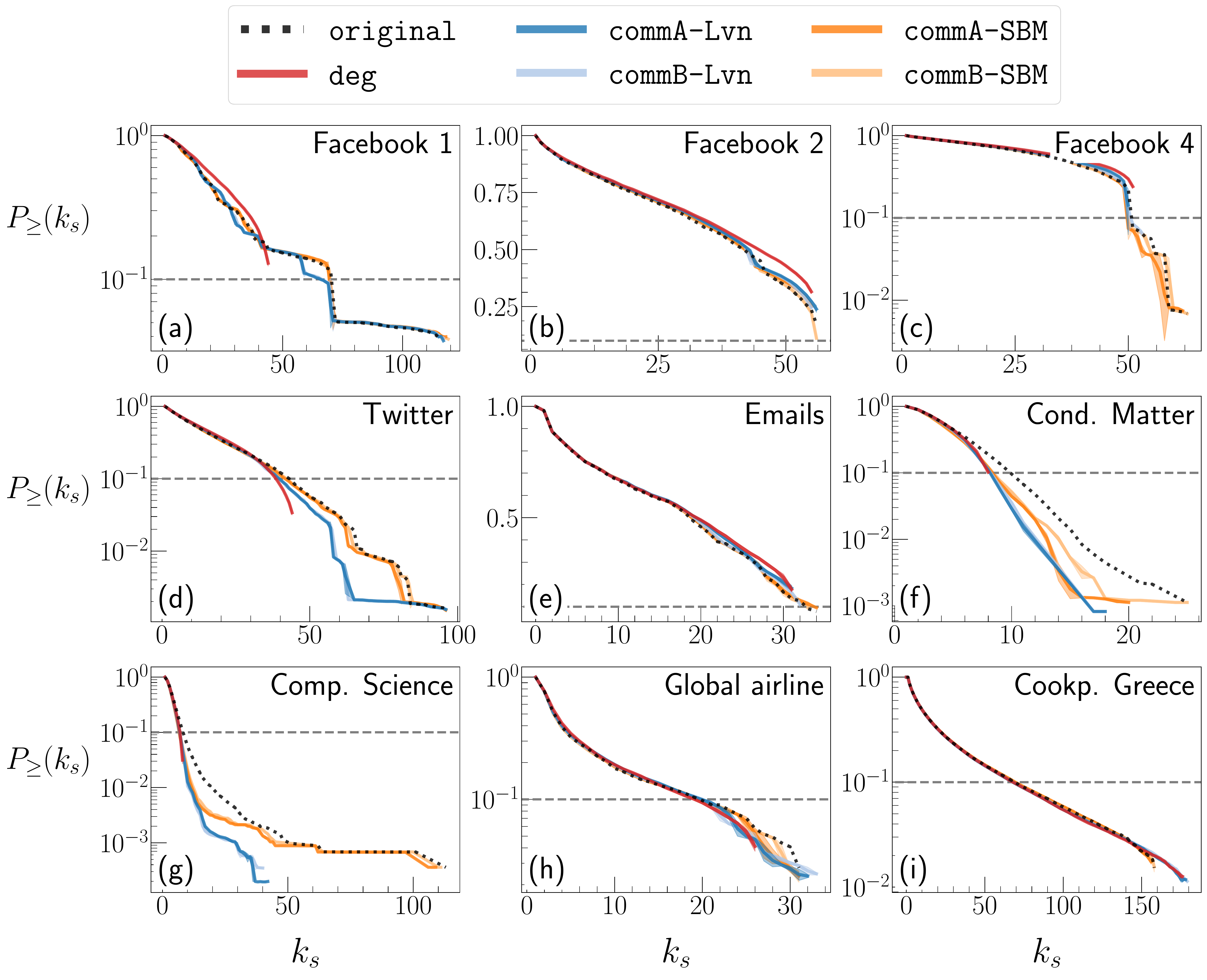}
\caption{Survival function of the probability distributions of the \kshell{} index, \ie{} $P_{\geq}(\ks)$ as a function of $\ks$ for the original network (dotted line) and shuffled networks (solid line). Each panel corresponds to a data set, \ie{} Facebook 1 (panel a), Facebook 2 (b), Facebook 4 (c), Twitter (d), Emails (e), Cond. Matter (f), Comp. Science (g), Global airline (h), and Cookpad Greece (i). The horizontal dashed lines indicate that $P_{\geq}(\ks) = 0.1$. Results are averaged over 10 different runs of each shuffling method, and the shaded areas (when visible) represent the standard deviations.}
\label{fig:cumR_all}
\end{figure}
%

%
%
%

%
%
\begin{figure}[ht]
\centering
\includegraphics[width=.85\linewidth]{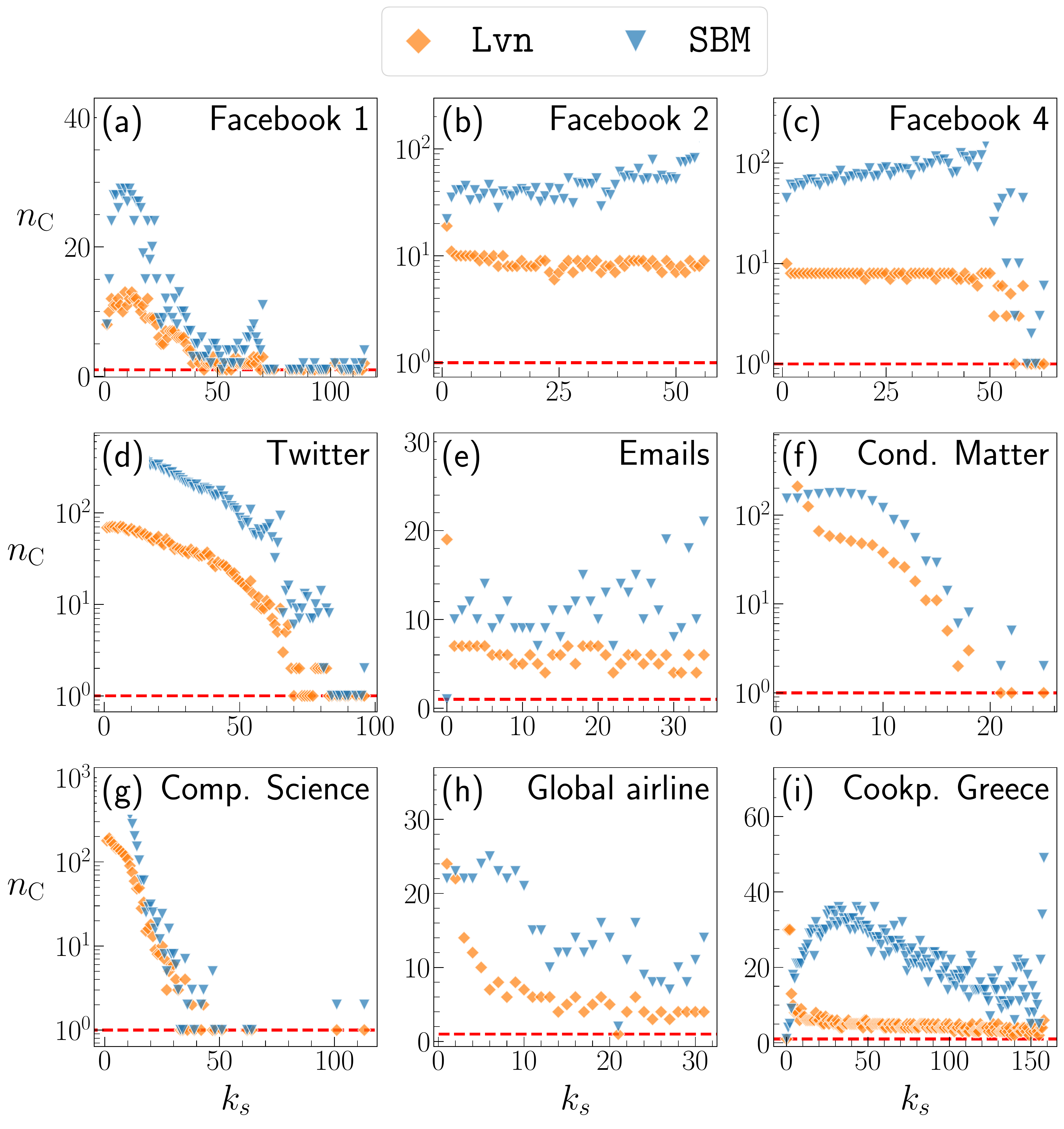}
\caption{Number of different communities, $\nc(\ks)$, that the set of nodes of a given \kshell{} value, $\ks$, overlaps. The horizontal dashed line is a guide to the eyes showing $\nc (\ks) = 1$. Each panel accounts for a different data set (see the caption of Fig.~\ref{fig:cumR_all} for the details). For each data set, we show the results corresponding to the community structure obtained using either \Louvain{} or \SBM{}.}
\label{fig:nrcomms_vs_kshell}
\end{figure}


%
%

\makeatletter

\makeatother

\end{document}


\title{Supplementary Materials for the manuscript entitled:\\Interplay between \kcore{} and community structure in complex networks}

\author{Irene Malvestio, Alessio Cardillo, \& Naoki Masuda}

\date{}


\setlength{\abovedisplayskip}{10pt}
\setlength{\belowdisplayskip}{10pt}

\maketitle

\section{Comparison between the original and shuffled networks}
\label{sec:network_comparison}

\begin{figure}[H]
\centering
\includegraphics[width=1\linewidth]{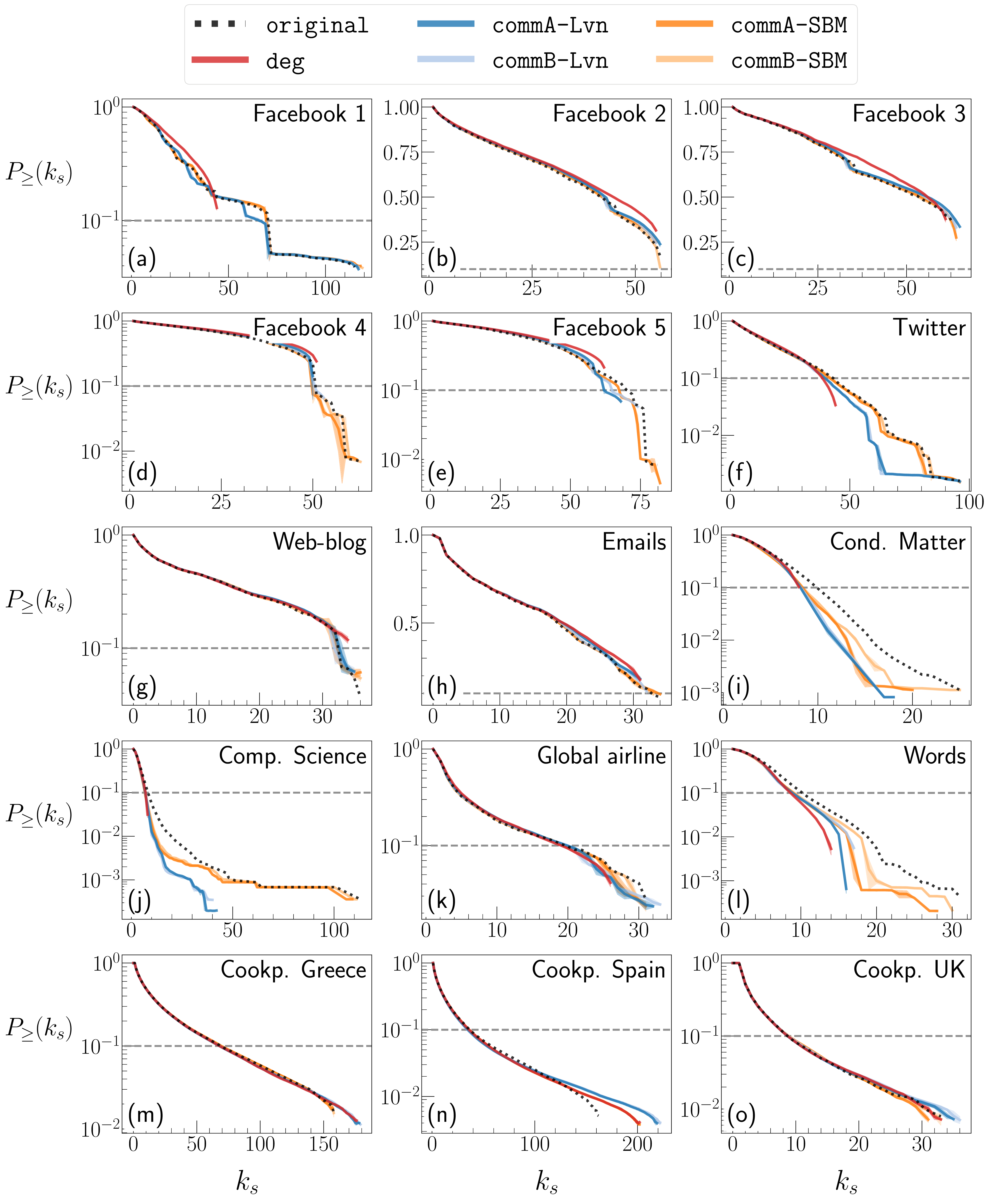}
\caption{Survival function of the probability distribution of the \kshell{} index, $P_{\geq}(\ks)$, as a function of $\ks$ for the empirical network (dotted lines) and shuffled networks (solid lines). Each panel corresponds to a data set. The horizontal dashed lines represent $P_{\geq}(\ks) = 0.1$. The results shown are averages over 10 different runs of each shuffling method, and the shaded areas (when visible) represent the standard deviations.}
\label{fig:cumR_other_data}
\end{figure}

In this section, we provide a detailed characterisation of the \kcore{} decomposition of the shuffled networks for all the empirical networks. Supplementary Figure~\ref{fig:cumR_other_data} shows the survival function of the probability distribution of the \kshell{} index, $P_{\geq}(\ks)$, of the original networks and their shuffled counterparts (\degswap{}, \commaswap{}, and \commbswap{}). The figure indicates that \commaswap{} and \commbswap{} produce \kshell{} distributions that are more similar to the original ones, compared to \degswap{}, in particular when \commaswap{} or \commbswap{} is combined with \SBM{}. This result also holds true for the Greece and Spain networks of Cookpad where, contrarily to the other data sets, the \degswap{} networks have a degeneracy, $D$, higher than the original networks.

\begin{figure}[H]
\centering
\includegraphics[width=.95\linewidth]{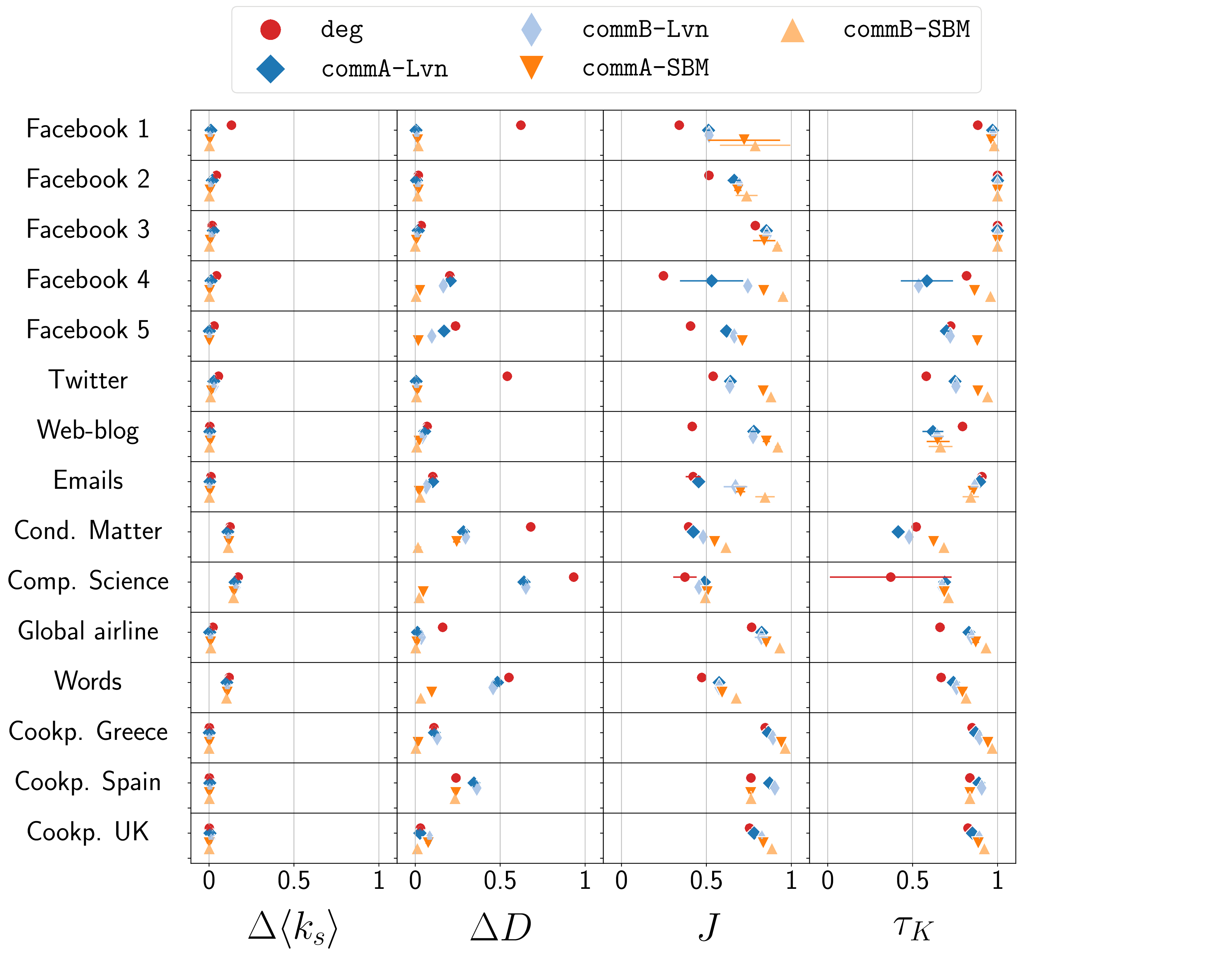}
\caption{Graphical summary of the values reported in Supplementary Table~\ref{tab:summary1}. For each pair of an empirical network and indicator, we show the value of the indicator for each shuffling method. The error bars represent the standard deviation.}
\label{fig:summary_difference}
\end{figure}

In Supplementary Table~\ref{tab:summary1}, we report the values of the four indicators used for comparing the \kcore{} decomposition between the original and shuffled networks. In particular, we report the average value and standard deviation of the relative difference $\Delta X = \left\lvert X(G) - X(G^\prime) \right\rvert/X(G)$ where $X$ is either the average \kshell{} index, $\avg{\ks}$, or $D$. In the same table, we also report the values of the Jaccard score, $J$, and Kendall's tau, $\tau_K$, calculated for the set of nodes belonging to the innermost \kshell{s} (see the main text for the details of the methods). We notice that, in general, \commbswap{-\SBM{}} yields the smallest values of $\Delta\avg{\ks}$ and $\Delta D$ and the largest values of $J$ and $\tau_K$; confirming its good performances in reconstructing the \kcore{} decomposition of the original network. Supplementary Figure~\ref{fig:summary_difference} provides an overview of the performances of each shuffling method. Figure~\ref{M-fig:summary_performance} in the main text is a projection of the information contained in Supplementary Fig.~\ref{fig:summary_difference}.

%
%
%

%
%
%
\begin{table}[H]
\centering
\begin{scriptsize}
\begin{tabular}{|c||l|c|c|c|c|c|c|c|}
\hline
\multicolumn{1}{|c||}{Data set} & \multicolumn{1}{|c|}{Indicator} & \multicolumn{1}{|c|}{\degswap} &  \multicolumn{1}{|c|}{\commaswap{--\Louvain{}}} & \multicolumn{1}{|c|}{\commbswap{--\Louvain{}}} &  \multicolumn{1}{|c|}{\commaswap{--\SBM{}}} & \multicolumn{1}{|c|}{\commbswap{--\SBM{}}} \\
\hline\hline
 \multirow{4}{*}{Facebook 1}
 & $\Delta \avg{\ks}$ & 0.132 $\pm$ 0.002 &0.011 $\pm$ 0.001 & 0.007 $\pm$ 0.003 & 0.004 $\pm$ 0.002 & 0.002 $\pm$ 0.001 \\ 
  & $\Delta D$ & 0.622 $\pm$ 0.004 &0.004 $\pm$ 0.006 & 0.009 $\pm$ 0.007 & 0.014 $\pm$ 0.006 & 0.017 $\pm$ 0.007 \\ 
 & $J$ & 0.340 $\pm$ 0.025 &0.512 $\pm$ 0.009 & 0.515 $\pm$ 0.004 & 0.722 $\pm$ 0.212 & 0.787 $\pm$ 0.207 \\ 
 & $\tau_K$ & 0.884 $\pm$ 0.005 &0.971 $\pm$ 0.014 & 0.974 $\pm$ 0.012 & 0.963 $\pm$ 0.015 & 0.980 $\pm$ 0.018 \\ 
\hline
 \multirow{4}{*}{Facebook 2} 
 & $\Delta \avg{\ks}$ & 0.043 $\pm$ 0.000 &0.020 $\pm$ 0.003 & 0.011 $\pm$ 0.001 & 0.006 $\pm$ 0.000 & 0.002 $\pm$ 0.001 \\ 
 & $\Delta D$ & 0.018 $\pm$ 0.000 &0.007 $\pm$ 0.009 & 0.018 $\pm$ 0.000 & 0.018 $\pm$ 0.000 & 0.012 $\pm$ 0.008 \\ 
 & $J$ & 0.515 $\pm$ 0.008 &0.664 $\pm$ 0.038 & 0.695 $\pm$ 0.014 & 0.685 $\pm$ 0.021 & 0.737 $\pm$ 0.064 \\ 
 & $\tau_K$ & 1.000 $\pm$ 0.000 &1.000 $\pm$ 0.000 & 1.000 $\pm$ 0.000 & 1.000 $\pm$ 0.000 & 1.000 $\pm$ 0.000 \\ 
\hline
 \multirow{4}{*}{Facebook 3} 
 & $\Delta \avg{\ks}$ & 0.019 $\pm$ 0.004 &0.025 $\pm$ 0.002 & 0.017 $\pm$ 0.003 & 0.007 $\pm$ 0.003 & 0.002 $\pm$ 0.001 \\ 
 & $\Delta D$ & 0.035 $\pm$ 0.006 &0.017 $\pm$ 0.005 & 0.011 $\pm$ 0.007 & 0.006 $\pm$ 0.008 & 0.000 $\pm$ 0.000 \\ 
 & $J$ & 0.788 $\pm$ 0.009 &0.853 $\pm$ 0.009 & 0.856 $\pm$ 0.027 & 0.840 $\pm$ 0.066 & 0.918 $\pm$ 0.017 \\ 
 & $\tau_K$ & 1.000 $\pm$ 0.000 &1.000 $\pm$ 0.000 & 1.000 $\pm$ 0.000 & 1.000 $\pm$ 0.000 & 1.000 $\pm$ 0.000 \\ 
\hline
 \multirow{4}{*}{Facebook 4} 
 & $\Delta \avg{\ks}$ & 0.044 $\pm$ 0.003 &0.012 $\pm$ 0.002 & 0.005 $\pm$ 0.002 & 0.003 $\pm$ 0.001 & 0.002 $\pm$ 0.001 \\ 
 & $\Delta D$ & 0.203 $\pm$ 0.006 &0.206 $\pm$ 0.000 & 0.165 $\pm$ 0.008 & 0.027 $\pm$ 0.010 & 0.005 $\pm$ 0.007 \\ 
 & $J$ & 0.247 $\pm$ 0.017 &0.531 $\pm$ 0.186 & 0.744 $\pm$ 0.015 & 0.837 $\pm$ 0.005 & 0.951 $\pm$ 0.016 \\ 
 & $\tau_K$ & 0.818 $\pm$ 0.011 &0.584 $\pm$ 0.154 & 0.535 $\pm$ 0.027 & 0.865 $\pm$ 0.008 & 0.959 $\pm$ 0.014 \\ 
\hline
 \multirow{4}{*}{Facebook 5} 
 & $\Delta \avg{\ks}$ & 0.030 $\pm$ 0.002 &0.002 $\pm$ 0.001 & 0.002 $\pm$ 0.001 & 0.001 $\pm$ 0.001 & --- \\ 
 & $\Delta D$ & 0.237 $\pm$ 0.005 &0.169 $\pm$ 0.006 & 0.098 $\pm$ 0.004 & 0.017 $\pm$ 0.006 & --- \\ 
 & $J$ & 0.406 $\pm$ 0.023 &0.619 $\pm$ 0.004 & 0.664 $\pm$ 0.004 & 0.712 $\pm$ 0.015 & --- \\ 
 & $\tau_K$ & 0.724 $\pm$ 0.016 &0.699 $\pm$ 0.016 & 0.721 $\pm$ 0.011 & 0.881 $\pm$ 0.014 & --- \\ 
\hline
 \multirow{4}{*}{Twitter}
 & $\Delta \avg{\ks}$ & 0.055 $\pm$ 0.000 &0.028 $\pm$ 0.000 & 0.030 $\pm$ 0.000 & 0.014 $\pm$ 0.000 & 0.008 $\pm$ 0.000 \\ 
  & $\Delta D$ & 0.542 $\pm$ 0.000 &0.005 $\pm$ 0.005 & 0.007 $\pm$ 0.005 & 0.011 $\pm$ 0.007 & 0.007 $\pm$ 0.005 \\ 
 & $J$ & 0.540 $\pm$ 0.001 &0.641 $\pm$ 0.008 & 0.637 $\pm$ 0.005 & 0.834 $\pm$ 0.003 & 0.881 $\pm$ 0.011 \\ 
 & $\tau_K$ & 0.580 $\pm$ 0.004 &0.749 $\pm$ 0.002 & 0.755 $\pm$ 0.002 & 0.884 $\pm$ 0.002 & 0.941 $\pm$ 0.002 \\ 
\hline
 \multirow{4}{*}{Web-blog}
 & $\Delta \avg{\ks}$ & 0.005 $\pm$ 0.002 &0.004 $\pm$ 0.003 & 0.004 $\pm$ 0.002 & 0.007 $\pm$ 0.004 & 0.003 $\pm$ 0.002 \\ 
  & $\Delta D$ & 0.069 $\pm$ 0.014 &0.056 $\pm$ 0.012 & 0.042 $\pm$ 0.014 & 0.022 $\pm$ 0.011 & 0.008 $\pm$ 0.013 \\ 
 & $J$ & 0.416 $\pm$ 0.009 &0.780 $\pm$ 0.021 & 0.775 $\pm$ 0.022 & 0.853 $\pm$ 0.021 & 0.920 $\pm$ 0.017 \\ 
 & $\tau_K$ & 0.794 $\pm$ 0.018 &0.619 $\pm$ 0.061 & 0.646 $\pm$ 0.038 & 0.651 $\pm$ 0.068 & 0.664 $\pm$ 0.070 \\ 
\hline
 \multirow{4}{*}{Emails} 
 & $\Delta \avg{\ks}$ & 0.011 $\pm$ 0.006 &0.005 $\pm$ 0.002 & 0.003 $\pm$ 0.001 & 0.005 $\pm$ 0.003 & 0.002 $\pm$ 0.001 \\ 
 & $\Delta D$ & 0.103 $\pm$ 0.015 &0.103 $\pm$ 0.015 & 0.065 $\pm$ 0.012 & 0.024 $\pm$ 0.012 & 0.029 $\pm$ 0.000 \\ 
 & $J$ & 0.422 $\pm$ 0.044 &0.454 $\pm$ 0.035 & 0.671 $\pm$ 0.069 & 0.701 $\pm$ 0.027 & 0.845 $\pm$ 0.057 \\ 
 & $\tau_K$ & 0.908 $\pm$ 0.008 &0.895 $\pm$ 0.008 & 0.864 $\pm$ 0.023 & 0.855 $\pm$ 0.011 & 0.842 $\pm$ 0.048 \\ 
\hline
 \multirow{4}{*}{Cond. Matter} 
 & $\Delta \avg{\ks}$ & 0.123 $\pm$ 0.001 &0.111 $\pm$ 0.001 & 0.113 $\pm$ 0.001 & 0.117 $\pm$ 0.001 & 0.113 $\pm$ 0.001 \\ 
 & $\Delta D$ & 0.680 $\pm$ 0.000 &0.284 $\pm$ 0.012 & 0.296 $\pm$ 0.020 & 0.244 $\pm$ 0.022 & 0.016 $\pm$ 0.020 \\ 
 & $J$ & 0.395 $\pm$ 0.006 &0.423 $\pm$ 0.013 & 0.481 $\pm$ 0.008 & 0.548 $\pm$ 0.013 & 0.615 $\pm$ 0.005 \\ 
 & $\tau_K$ & 0.521 $\pm$ 0.018 &0.415 $\pm$ 0.025 & 0.479 $\pm$ 0.016 & 0.623 $\pm$ 0.013 & 0.685 $\pm$ 0.008 \\ 
\hline
 \multirow{4}{*}{Comp. Science} 
 & $\Delta \avg{\ks}$ & 0.171 $\pm$ 0.004 &0.153 $\pm$ 0.000 & 0.159 $\pm$ 0.000 & 0.146 $\pm$ 0.000 & 0.144 $\pm$ 0.000 \\ 
 & $\Delta D$ & 0.933 $\pm$ 0.004 &0.641 $\pm$ 0.006 & 0.651 $\pm$ 0.004 & 0.047 $\pm$ 0.007 & 0.023 $\pm$ 0.004 \\ 
 & $J$ & 0.374 $\pm$ 0.069 &0.488 $\pm$ 0.002 & 0.456 $\pm$ 0.002 & 0.504 $\pm$ 0.003 & 0.493 $\pm$ 0.002 \\ 
 & $\tau_K$ & 0.371 $\pm$ 0.358 &0.688 $\pm$ 0.003 & 0.673 $\pm$ 0.002 & 0.686 $\pm$ 0.002 & 0.711 $\pm$ 0.002 \\ 
\hline
 \multirow{4}{*}{Global airline} 
  & $\Delta \avg{\ks}$ & 0.023 $\pm$ 0.001 &0.004 $\pm$ 0.003 & 0.008 $\pm$ 0.003 & 0.009 $\pm$ 0.003 & 0.010 $\pm$ 0.002 \\ 
 & $\Delta D$ & 0.161 $\pm$ 0.000 &0.013 $\pm$ 0.016 & 0.035 $\pm$ 0.017 & 0.010 $\pm$ 0.015 & 0.003 $\pm$ 0.010 \\ 
 & $J$ & 0.766 $\pm$ 0.011 &0.826 $\pm$ 0.010 & 0.824 $\pm$ 0.039 & 0.852 $\pm$ 0.016 & 0.933 $\pm$ 0.008 \\ 
 & $\tau_K$ & 0.661 $\pm$ 0.016 &0.831 $\pm$ 0.006 & 0.846 $\pm$ 0.014 & 0.872 $\pm$ 0.020 & 0.932 $\pm$ 0.006 \\ 
\hline
 \multirow{4}{*}{Words} 
 & $\Delta \avg{\ks}$ & 0.118 $\pm$ 0.000 &0.105 $\pm$ 0.000 & 0.108 $\pm$ 0.000 & 0.107 $\pm$ 0.000 & 0.102 $\pm$ 0.000 \\ 
 & $\Delta D$ & 0.552 $\pm$ 0.010 &0.484 $\pm$ 0.000 & 0.458 $\pm$ 0.013 & 0.097 $\pm$ 0.000 & 0.032 $\pm$ 0.000 \\ 
 & $J$ & 0.473 $\pm$ 0.003 &0.575 $\pm$ 0.001 & 0.575 $\pm$ 0.002 & 0.592 $\pm$ 0.001 & 0.675 $\pm$ 0.002 \\ 
 & $\tau_K$ & 0.668 $\pm$ 0.003 &0.740 $\pm$ 0.001 & 0.758 $\pm$ 0.003 & 0.794 $\pm$ 0.003 & 0.815 $\pm$ 0.001 \\ 
\hline
 \multirow{4}{*}{Cookpad -- Greece}
 & $\Delta \avg{\ks}$ & 0.002 $\pm$ 0.000 &0.001 $\pm$ 0.000 & 0.002 $\pm$ 0.000 & 0.002 $\pm$ 0.001 & 0.000 $\pm$ 0.000 \\ 
 & $\Delta D$ & 0.110 $\pm$ 0.003 &0.113 $\pm$ 0.007 & 0.129 $\pm$ 0.003 & 0.016 $\pm$ 0.006 & 0.004 $\pm$ 0.003 \\ 
 & $J$ & 0.846 $\pm$ 0.003 &0.865 $\pm$ 0.003 & 0.892 $\pm$ 0.004 & 0.941 $\pm$ 0.002 & 0.964 $\pm$ 0.002 \\ 
 & $\tau_K$ & 0.850 $\pm$ 0.001 &0.871 $\pm$ 0.001 & 0.894 $\pm$ 0.001 & 0.943 $\pm$ 0.002 & 0.968 $\pm$ 0.004 \\ 
\hline
 \multirow{4}{*}{Cookpad -- Spain}
 & $\Delta \avg{\ks}$ & 0.002 $\pm$ 0.000 &0.005 $\pm$ 0.000 & 0.005 $\pm$ 0.000 & 0.002 $\pm$ 0.000 & 0.002 $\pm$ 0.000 \\ 
  & $\Delta D$ & 0.240 $\pm$ 0.006 &0.346 $\pm$ 0.006 & 0.362 $\pm$ 0.005 & 0.238 $\pm$ 0.005 & 0.234 $\pm$ 0.006 \\ 
 & $J$ & 0.762 $\pm$ 0.006 &0.872 $\pm$ 0.004 & 0.903 $\pm$ 0.003 & 0.760 $\pm$ 0.005 & 0.762 $\pm$ 0.006 \\ 
 & $\tau_K$ & 0.837 $\pm$ 0.001 &0.890 $\pm$ 0.001 & 0.907 $\pm$ 0.000 & 0.837 $\pm$ 0.001 & 0.837 $\pm$ 0.001 \\ 
\hline
 \multirow{4}{*}{Cookpad -- UK} & 
 $\Delta \avg{\ks}$ & 0.001 $\pm$ 0.001 &0.005 $\pm$ 0.001 & 0.012 $\pm$ 0.001 & 0.001 $\pm$ 0.001 & 0.001 $\pm$ 0.000 \\ 
 &  $\Delta D$ & 0.030 $\pm$ 0.014 &0.027 $\pm$ 0.016 & 0.085 $\pm$ 0.012 & 0.076 $\pm$ 0.015 & 0.012 $\pm$ 0.015 \\ 
 & $J$ & 0.754 $\pm$ 0.010 &0.781 $\pm$ 0.007 & 0.827 $\pm$ 0.005 & 0.835 $\pm$ 0.003 & 0.886 $\pm$ 0.006 \\ 
 & $\tau_K$ & 0.826 $\pm$ 0.003 &0.852 $\pm$ 0.003 & 0.893 $\pm$ 0.002 & 0.886 $\pm$ 0.002 & 0.922 $\pm$ 0.002 \\ 
\hline
\end{tabular}
\end{scriptsize}
\caption{Average and standard deviation of the four indicators characterising the \kcore{} decomposition. In the cells with missing values, the shuffling method did not converge.}
\label{tab:summary1} 
\end{table}
%

%
%
%

%
%

\newpage

\section{Effects of changing the resolution of the Louvain method}
\label{sec:louvain_resolution} 

A pitfall of the Louvain method is its propensity to merge small communities due to the existence of a lower bound in the size of the communities (\ie{} the modularity's resolution limit)~\cite{fortunato2007resolution,fortunato2016community}. However, communities of empirical networks may not have a typical size, and small communities coexist with large ones in general. A method for mitigating the resolution limit is to introduce a resolution parameter $r \in (0,1]$ into the Louvain algorithm~\cite{lambiotte-ieee_trans_netsci_eng-2014}. By tuning $r$, it is possible to vary the resolution scale of the detected communities, spanning from large (\ie{} $r \sim 1$) to small (\ie{} $r \sim 0$) communities. We denote the Louvain method using a different resolution $r$ by \LouvainR{}.

Figure~\ref{M-fig:summary_performance} in the main text shows that \commbswap{} combined with the communities found by the \SBM{} reproduces the tail of $P_{\geq}(\ks)$ of the empirical networks most accurately. Table~\ref{M-tab:data} indicates that \SBM{} tends to find more communities than \Louvain{}, although there are exceptions. To understand whether the different number of communities found by \SBM{} and \Louvain{} is the reason behind their different performances, we extracted the communities using \LouvainR{} with $r \in \{ 0.1, 0.3, 0.5 \}$. For the sake of brevity, we only discuss the results for $r = 0.3$ in the following text. However, we have verified that the results for the other values of $r$ are similar.

\begin{figure}[H]
\centering
\includegraphics[width=\linewidth]{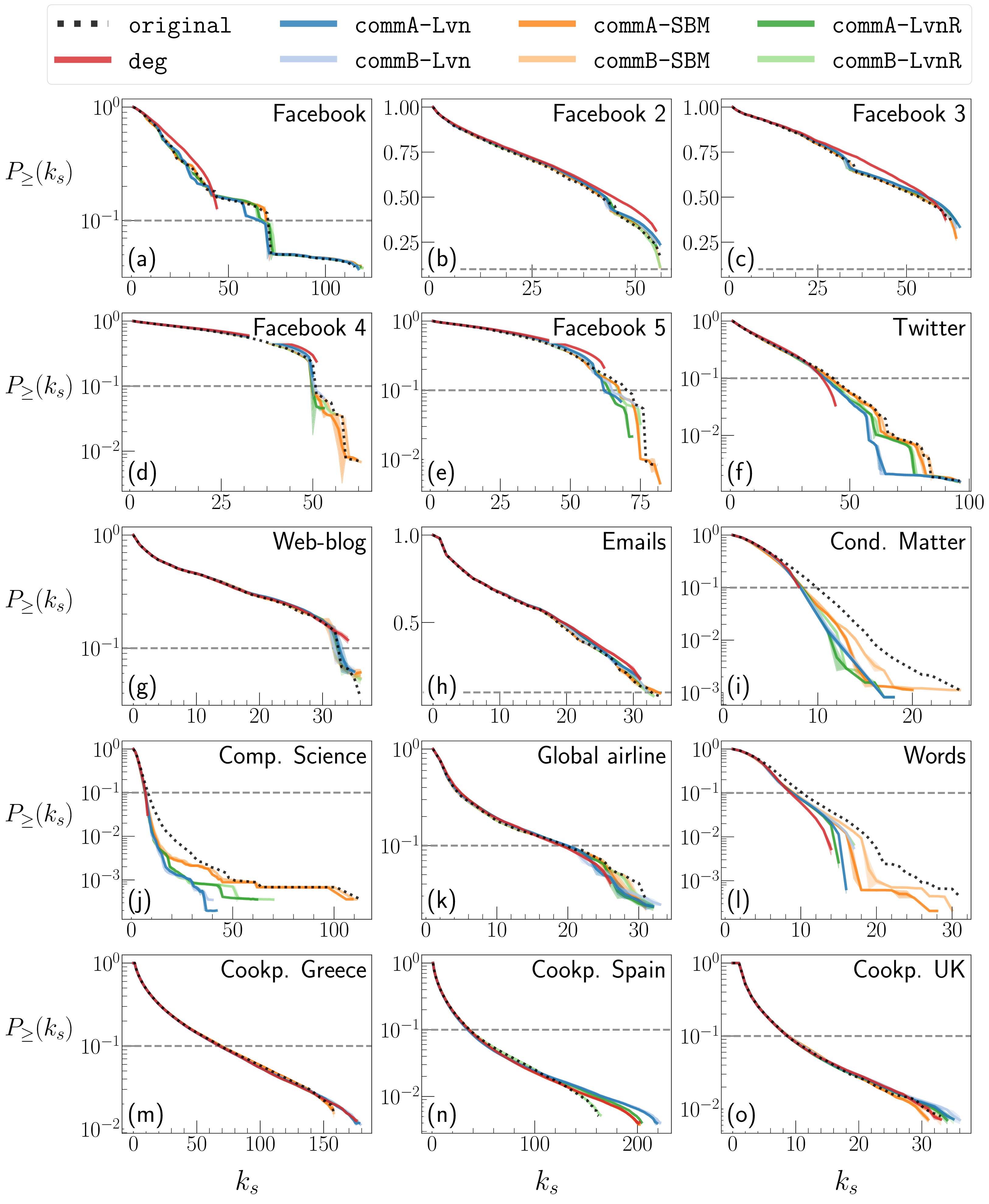}
\caption{Survival function of the probability distribution of the \kshell{} index, $P_{\geq}(\ks)$, as a function of $\ks$ for all the cases considered in Fig.~\ref{M-fig:cumR_all} plus the cases in which we identify communities using the Louvain method with a resolution parameter of $r = 0.3$. See the caption or Supplementary Fig.~\ref{fig:cumR_other_data} for notations and legends.}
\label{fig:cumR_Lr3}
\end{figure}

\begin{figure}[H]
\centering
\includegraphics[width=\linewidth]{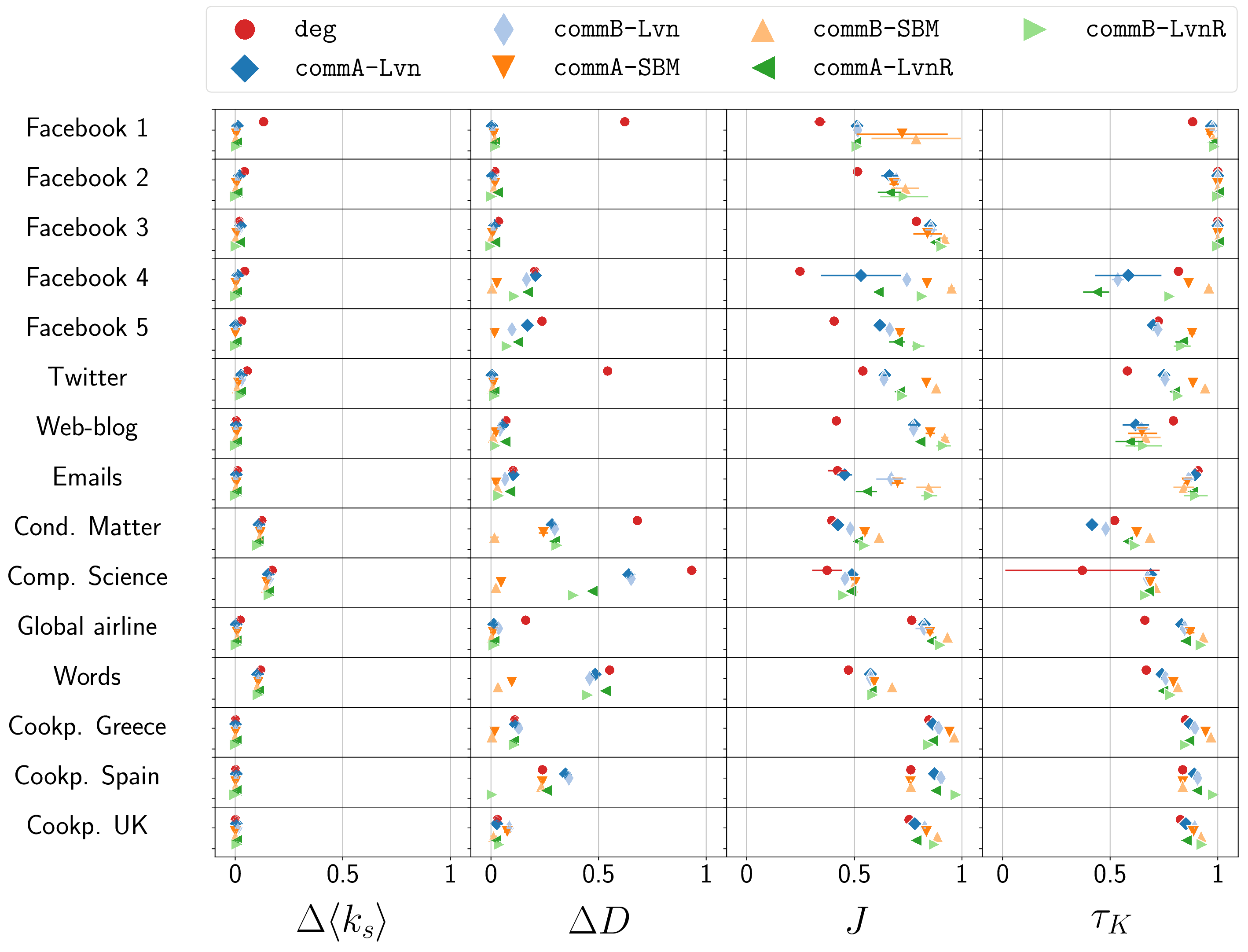}
\caption{Graphical summary of the average and standard deviation of the four indicators characterising the \kcore{} decomposition. In addition to the information displayed in Supplementary Fig.~\ref{fig:summary_difference}, we consider the community structure detected by the Louvain method with $r = 0.3$ (\LouvainR{}). See the caption or Supplementary Fig.~\ref{fig:summary_difference} for notations and legends.}
\label{fig:summary_difference_Lr}
\end{figure}

In Supplementary Table~\ref{tab:data_r} we show, for all the data sets considered, the number of communities, $\NC$, and the modularity, $Q$, corresponding to the community structure found using \Louvain{}, \SBM{}, and \LouvainR{} with $r = 0.3$. With \LouvainR{}, the number of communities is similar or even larger than that obtained with \SBM{}. Supplementary Figures~\ref{fig:cumR_Lr3} and \ref{fig:summary_difference_Lr} show $P_{\geq}(\ks)$ plotted against $\ks$ and the four indicators used for comparing the innermost \kshell{s}, respectively, including the results obtained with \LouvainR{}. These figures indicate that using $r=0.3$ as opposed to $r=1$ improves the ability of the Louvain algorithm to mimic the structure of the \kshell{}. In Supplementary Fig.~\ref{fig:summary_performance_Lr} we summarise the performances of the different shuffling methods including \LouvainR{}. 

Although \LouvainR{} mimics the \kshell{} features better than \SBM{} for some data sets and indicators (approximately around the $20\%$ of the cases), \SBM{} still attains the highest success ratio for each indicator, $f_X$. The few cases for which \commbswap{-\LouvainR{}} does better than \commbswap{-\SBM{}} are the networks for which the difference between the empirical $P_{\geq}(\ks)$ and $P_{\geq}(\ks)$ obtained from the \degswap{} shuffling apparently looks small, such as the Emails and Cookpad UK networks. In these networks, the differences between the $P_{\geq}(\ks)$ of the networks obtained using \commbswap{-\LouvainR{}} and \commbswap{-\SBM{}} are also small. 
By contrast, other data sets such as Condensed Matter, Computer Science, and Words show bigger differences between their $P_{\geq}(\ks)$ and that obtained using the \degswap{} method. In these data sets, \commbswap{-\SBM{}} is much better than \commbswap{-\LouvainR{}}, although \commbswap{-\LouvainR{}} is better than \commbswap{-\Louvain{}}. Overall, these results suggest that the increase in the number of communities enabled by a small $r$ does not lead to reconstruction of the \kshell{} structure with a better accuracy than \SBM{} does.

%
%
\begin{table}[H]
\centering
\newcolumntype{d}[1]{D{.}{.}{#1} }
\begin{tabular}{|l|cc|cc|cc|}
 \hline 
Data set & $\NC^{\text{\Louvain{}}}$ & $Q^\text{\Louvain{}}$ & $\NC^{\text{\SBM{}}}$ & $Q^\text{\SBM{}}$ & $\NC^{\text{\LouvainR{}}}$ & $Q^\text{\LouvainR{}}$ \\ \hline\hline
Facebook 1     &  16  & 0.835 & 62  & 0.551 & 29  & 0.819 \\
Facebook 2     &  19  & 0.419 & 198 & 0.158 & 67  & 0.348 \\
Facebook 3     &  8   & 0.436 & 87  & 0.139 & 36  & 0.294 \\
Facebook 4     &  10  & 0.438 & 274 & 0.193 & 72  & 0.381 \\
Facebook 5     &  18  & 0.470 & 547 & 0.172 & 92  & 0.417 \\
Twitter        &  73  & 0.808 & 510 & 0.511 & 136 & 0.779 \\
Web-blogs      &  275 & 0.426 & 17  & 0.076 & 331 & 0.150 \\
Emails         &  26  & 0.410 & 33  & 0.232 & 61  & 0.350 \\
Cond. Matter   &  619 & 0.730 & 203 & 0.633 & 716 & 0.718 \\
Comp. Science  &  209 & 0.822 & 676 & 0.726 & 438 & 0.812 \\
Global airline &  26  & 0.665 & 40  & 0.311 & 55  & 0.542 \\
Words          &  378 & 0.759 & 548 & 0.583 & 523 & 0.747 \\
Cookpad Greece &  40  & 0.166 & 76  & 0.020 & 365 & 0.067 \\
Cookpad Spain  &  262 & 0.270 & 90  & 0.035 & 501 & 0.164 \\
Cookpad UK     &  199 & 0.350 & 8   & 0.114 & 320 & 0.314 \\ \hline
\end{tabular}
\caption{Summary of the properties of the community structures of the data sets analysed using different community detection methods. For each pair (network, method) we computed the number of communities, $\NC$, and the value of the modularity $Q$. Each pair of columns accounts for a different method, namely: Louvain (\Louvain{}), stochastic block model (\SBM{}), and Louvain with a resolution parameter $r = 0.3$ (\LouvainR{}).}
\label{tab:data_r} 
\end{table}
%

%
%
%

%
%
%

%
%
\begin{figure}[H]
\centering
\includegraphics[width=\linewidth]{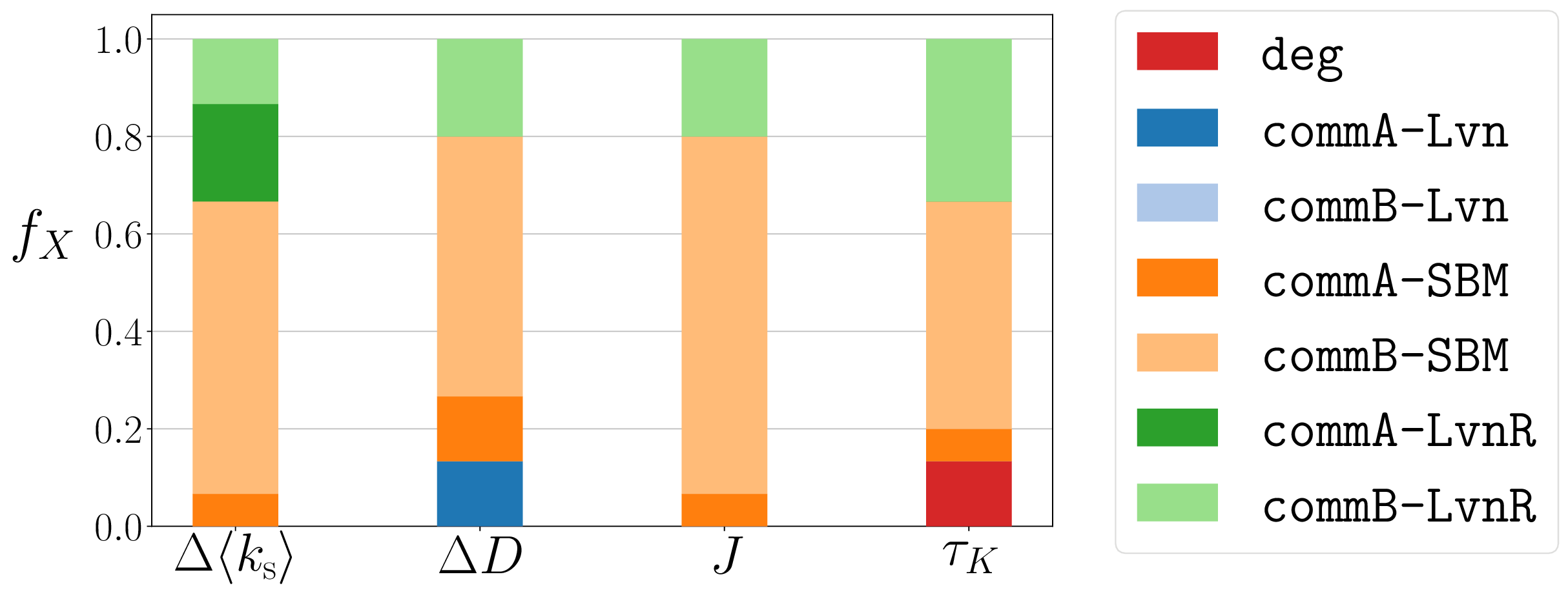}
\caption{Summary of the performances of the different shuffling methods in reproducing the features of the \kshell{s} according to four indicators. We report the fraction of data sets for which a given combination of the shuffling method and the community detection method yields an indicator's value closest to that for the original network. In addition to the methods considered in Fig.~\ref{M-fig:summary_performance}, we also consider the case of the communities extracted using the Louvain method with $r = 0.3$. See the caption or Fig.~\ref{M-fig:summary_performance} for notations and legends.}
\label{fig:summary_performance_Lr}
\end{figure}
%

To investigate which features of the SBM and Louvain algorithms are responsible for the difference in their performances, we study the average of the size of the communities to which nodes of a certain \kshell{} belong, $\mcs$, as a function of the \kshell{} index, $\ks$. In Supplementary Fig.~\ref{fig:mean_size_com}, we observe how $\mcs$ of the communities found by the Louvain method (\Louvain{}) stays nearly constant across the entire range of $\ks$ values. By contrast, with \SBM{}, $\mcs$ monotonically decreases as $\ks$ increases, such that nodes in inner \kshell{s} tend to belong to smaller communities. Although a high resolution (\ie{} a small $r$ value) in the Louvain method produces a large number of communities (see Supplementary Table~\ref{tab:data_r}), the behaviour of $\mcs$ for \Louvain{} and \LouvainR{} is similar, with the major difference that the value of $\mcs$ is smaller for \LouvainR{}.

These results altogether lead us to conclude that our method combined with the community structure identified using the Louvain method with a higher resolution does not outperform our method combined with \SBM{} in grasping the features of the \kcore{} decomposition. This may be because \SBM{} is capable of finding more universal mesostructures than those found by the Louvain method~\cite{olhede-pnas-2014,newman-pre-2016,young-pre-2018}.

%
%
\begin{figure}[H]
\centering
\includegraphics[width=.85\linewidth]{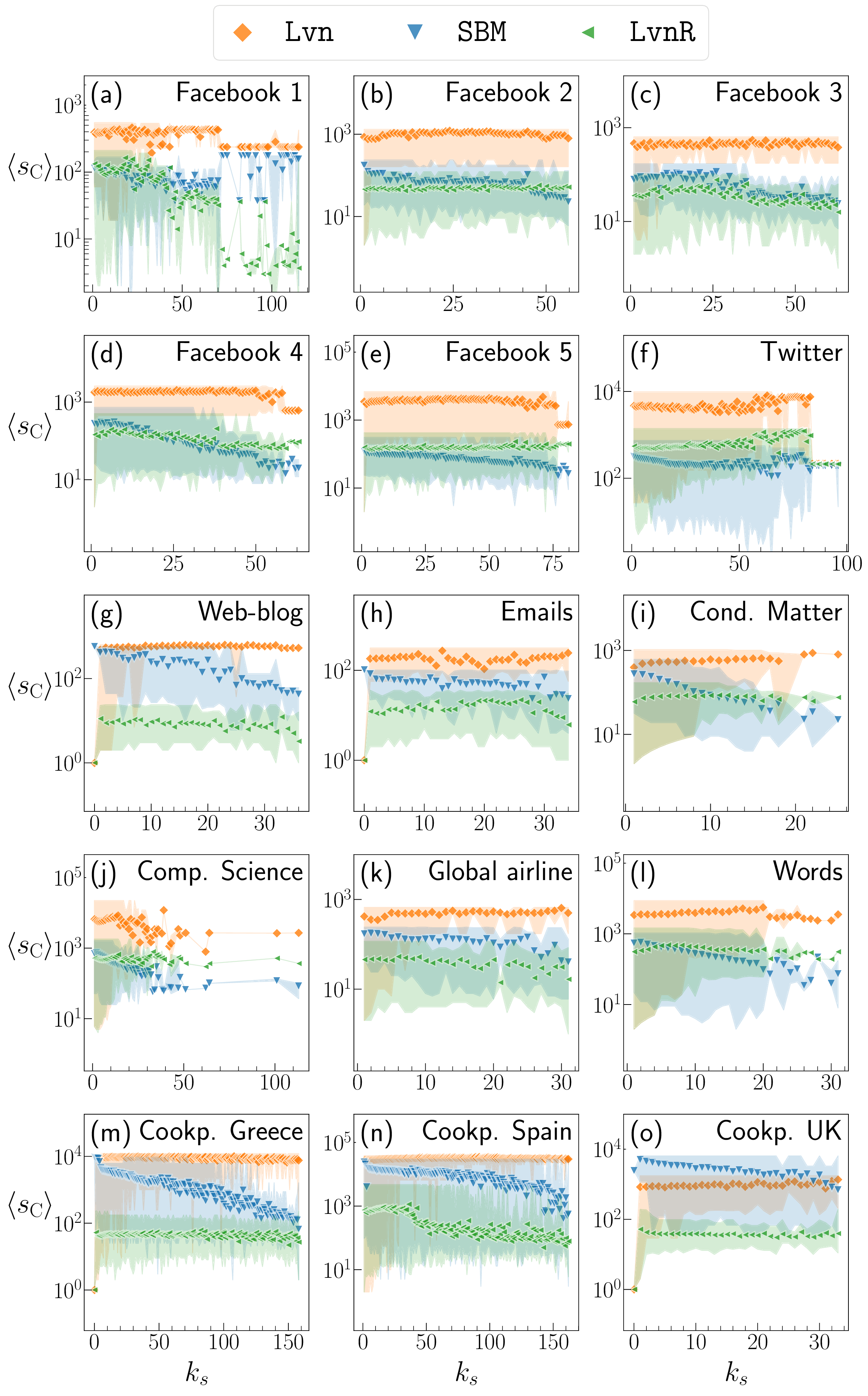}
\caption{Average size of the communities to which the nodes having \kshell{} index $\ks$ belong to, $\langle s_{\text{C}} \rangle$, versus $\ks$. We identify the community structure using either the Louvain (\Louvain{}), the stochastic block model (\SBM{}), or the Louvain with higher resolution (\LouvainR{}) methods. The resolution parameter for (\LouvainR{}) is equal to $r = 0.3$. Shaded areas denote the standard deviation. Each panel accounts for a different data set.}
\label{fig:mean_size_com}
\end{figure}
%

\newpage


\section{The LFR model}
\label{sec:lfr}

The Lancichinetti-Fortunato-Radicchi (LFR) model generates networks where both the node's degree and the size of the communities (\ie{} the number of nodes belonging to a community) follow power-law distributions \cite{lancichinetti-pre-2008}. Such features are found in many empirical networks \cite{amaral-pnas-2000} and have led to the success of the LFR model as generator of benchmark networks to test community detection algorithms \cite{fortunato2016community}. A main finding presented in the main text is that preserving the community structure of the original network in addition to the degree of each node improves the ability of the shuffling methods to mimic the \kcore{} decomposition of the original networks. Here, to test whether or not the community structure and the degree of each node, but not a possible intricate association between the two, is sufficient for mimicking the features of \kcore{} decomposition observed for many empirical networks, we generated networks using the LFR model and analysed their \kcore{s} and those of the shuffled counterparts.

The LFR algorithm depends on the following parameters: the exponent, $t_1 \in [2,3]$, of the degree distribution $P(k) \propto k^{-t_1}$; the exponent, $t_2 \in [1,2]$, of the community's size distribution $P(S_c) \propto {S_c}^{-t_2}$; the mixing parameter, $\mu \in [0,1]$, specifying the fraction of intra-community edges for a node. A value of $\mu = 0$ indicates that a node is connected only with nodes belonging to communities different from its own. A value of $\mu = 1$ indicates that a node is connected exclusively with nodes belonging to its own community; either one of the following: the average degree, $\avg{k}$, the minimum degree, $k_{\min}$, or the minimum number of communities, $\min\NC$. This stochastic algorithm may not produce a network fulfilling all the requirements in some realisations. Therefore, we have to set the parameter values to ensure the algorithm's convergence.

To encompass a good spectrum of networks, we consider four batches of parameter sets, which are summarised in Supplementary Table~\ref{tab:lfr_summary}, together with the properties of the generated networks. Each batch of parameter sets consists of a value of $t_1$, a value of $t_2$, and seven values of $\mu$ ranging from $0.1$ to $0.8$. We assumed $N=10000$ nodes and used the implementation of the LFR algorithm in the NetworkX Python package \cite{hagberg-scipy-2008}.

For each network generated, we extracted its \kcore{} decomposition and calculated the four indicators. We did the same for the shuffled counterparts generated using the \degswap{}, \commaswap{}, and \commbswap{} methods. In analogy to Supplementary Fig.~\ref{fig:cumR_other_data}, in Supplementary Figs.~\ref{fig:cumR_LFR_G1}--\ref{fig:cumR_LFR_G4} we show the survival function of the probability distribution of the \kshell{} index, $P_{\geq}(\ks)$, for the original LFR networks and the shuffled counterparts, one figure per each $(t_1, t_2)$ pair. An eye inspection of Supplementary Figs.~\ref{fig:cumR_LFR_G1}--\ref{fig:cumR_LFR_G4} highlights the existence of three trends. 

First, Supplementary Figs.~\ref{fig:cumR_LFR_G1} and \ref{fig:cumR_LFR_G2} indicate that, in networks generated using the smaller $t_1$ values (\ie{} parameters batches 1 and 2 in Supplementary Table~\ref{tab:lfr_summary}), the shuffled networks generated by \degswap{}, \commaswap{-\Louvain{}}, and \commbswap{-\Louvain{}} attain a \kcore{} decomposition with a degeneracy, $D$, considerably higher than the original one. In contrast, Supplementary Figs.~\ref{fig:cumR_LFR_G3} and \ref{fig:cumR_LFR_G4} indicate that, with the larger $t_1$ values (\ie{} parameter batches 3 and 4), we recover the same trend as that shown in Fig.~\ref{M-fig:cumR_all}. In other words, $D$ for the original networks are larger than that for the shuffled networks. The difference between the original $D$ and its shuffled counterpart seems to be influenced by the value of $t_1$, but not $t_2$ or $\mu$. 

Second, $P_{\geq}(\ks)$ for the original LFR networks mainly decreases smoothly as $\ks$ increases, without plateaus or abrupt drops. Therefore, the \kcore{} decomposition of LFR networks does not return any \kshell{} that is empty or much more populated than its adjacent \kshell{s}. This result is in stark contrast to that for various empirical networks, \eg{} the Facebook 1 data set (see Supplementary Fig.~\ref{fig:cumR_other_data}).

Third, regardless of the values of $t_1$, $t_2$, and $\mu$, the \commbswap{-\SBM{}} shuffling method produces networks with the $P_{\geq}(\ks)$ more akin to the original one than the other shuffling methods do. This result is consistent with that for the empirical networks presented in the main text.

In a nutshell, the analysis of the \kcore{} decomposition of networks generated by the LFR model reveals that the presence of communities is not enough to justify main properties of the \kshell{} structure observed in the empirical networks.

%
%
\begin{table}[H]
\centering
\begin{small}
\begin{tabular}{|c|c||cccccccccc|}
\hline 
\multicolumn{2}{|c||}{LFR parameters} & \multirow{2}{*}{$L$} & \multirow{2}{*}{$k_{\min}$} & \multirow{2}{*}{$\avg{k}$} & \multirow{2}{*}{$k_{\max}$} & \multirow{2}{*}{$\avg{\ks}$} & \multirow{2}{*}{$D$} & \multirow{2}{*}{$\NC^{\text{\Louvain{}}}$} & \multirow{2}{*}{$Q^\text{\Louvain{}}$} & \multirow{2}{*}{$\NC^{\text{\SBM{}}}$} & \multirow{2}{*}{$Q^\text{\SBM{}}$}\\\cline{1-2} 
\centering $t_1$, $t_2$ & $\mu$ &  &  &  &  &  &  &  &  & & \\\hline
%
\multicolumn{12}{c}{\textbf{Parameter batch 1}}\\\hline
\multirow{7}{*}{$\begin{array}{ccc} t_1 & = & 2.2\\ t_2 & = & 1.5\end{array}$}
& 0.1 & 120893 & 4 & 24.179 & 3470 & 12.541 & 16 & 10 & 0.673 & 29 & 0.041  \\
& 0.2 & 116200 & 4 & 23.240 & 3380 & 12.201 & 14 & 4  & 0.497 & 16 & -0.006 \\
& 0.3 & 118260 & 4 & 23.652 & 3199 & 12.188 & 14 & 7  & 0.458 & 26 & -0.002 \\
& 0.4 & 118547 & 4 & 23.709 & 6309 & 12.287 & 14 & 7  & 0.234 & 19 & -0.053 \\
& 0.5 & 130304 & 4 & 26.061 & 4481 & 12.548 & 16 & 7  & 0.250 & 24 & -0.038 \\ 
& 0.6 & 126277 & 4 & 25.255 & 4607 & 12.967 & 15 & 10 & 0.164 & 8  & -0.151 \\ 
& 0.8 & 118032 & 4 & 23.606 & 4028 & 12.263 & 14 & 10 & 0.162 & 5  & -0.156 \\\hline 
%
\multicolumn{12}{c}{\textbf{Parameter batch 2}}\\\hline
\multirow{7}{*}{$\begin{array}{ccc} t_1 & = & 2.6\\ t_2 & = & 2.0\end{array}$}
& 0.1 & 132920 & 8 & 26.584 & 2474 & 14.277 & 16 & 5 & 0.651 & 16 &  0.118  \\
& 0.2 & 129069 & 8 & 25.814 & 1641 & 14.186 & 15 & 8 & 0.595 & 22 &  0.139  \\
& 0.3 & 129024 & 8 & 25.805 & 1287 & 14.138 & 15 & 9 & 0.487 & 26 &  0.091  \\
& 0.4 & 128606 & 8 & 25.721 & 3305 & 14.015 & 15 & 7 & 0.222 & 9  &  -0.020 \\
& 0.5 & 127596 & 8 & 25.519 & 1504 & 14.105 & 15 & 8 & 0.227 & 20 &  0.041  \\
& 0.6 & 131024 & 8 & 26.205 & 1287 & 14.165 & 15 & 7 & 0.178 & 5  &  -0.093 \\
& 0.8 & 133017 & 8 & 26.603 & 4436 & 14.665 & 16 & 8 & 0.166 & 8  &  -0.087 \\\hline
%
\multicolumn{12}{c}{\textbf{Parameter batch 3}}\\\hline
\multirow{7}{*}{$\begin{array}{ccc} t_1 & = & 2.9\\ t_2 & = & 1.5\end{array}$}
& 0.1 & 315105 & 24 & 63.012 & 4249 &  35.907 & 37 & 4  & 0.511  & 14 &  0.115  \\
& 0.2 & 320836 & 24 & 64.167 & 2439 &  36.362 & 37 & 12 & 0.584  & 31 &  0.179  \\
& 0.3 & 319482 & 24 & 63.896 & 3732 &  36.234 & 37 & 10 & 0.359  & 29 &  0.078  \\
& 0.4 & 319070 & 24 & 63.814 & 2371 &  36.102 & 37 & 11 & 0.314  & 30 &  0.063  \\
& 0.5 & 321222 & 24 & 64.244 & 2795 &  36.816 & 38 & 9  & 0.204  & 27 &  0.029  \\
& 0.6 & 317738 & 24 & 63.548 & 2371 &  36.019 & 37 & 8  & 0.146  & 18 &  0.016  \\
& 0.8 & 305945 & 24 & 61.189 & 2246 &  35.049 & 36 & 9  & 0.109  & 4  &  -0.052 \\\hline
%
\multicolumn{12}{c}{\textbf{Parameter batch 4}}\\\hline
\multirow{7}{*}{$\begin{array}{ccc} t_1 & = & 3.0\\ t_2 & = & 2.0\end{array}$}
& 0.1 & 247311 & 20 & 49.462 & 2819 & 28.561 & 31 & 31 &  0.740 & 57 & 0.323  \\
& 0.2 & 246506 & 20 & 49.301 & 1228 & 27.691 & 28 & 40 &  0.651 & 69 & 0.331  \\
& 0.3 & 254822 & 20 & 50.964 & 1779 & 29.468 & 30 & 29 &  0.484 & 50 & 0.224  \\
& 0.4 & 249528 & 20 & 49.906 & 1131 & 28.457 & 29 & 37 &  0.387 & 71 & 0.156  \\
& 0.5 & 254371 & 20 & 50.874 & 2668 & 29.120 & 30 & 17 &  0.211 & 42 & 0.070  \\
& 0.6 & 243746 & 20 & 48.749 & 1097 & 28.311 & 29 & 20 &  0.186 & 50 & 0.073  \\
& 0.8 & 251094 & 20 & 50.219 & 3569 & 28.364 & 29 & 9  &  0.119 & 4  & -0.053 \\\hline
\end{tabular}
\end{small}
\caption{Summary of the properties of the networks generated with the LFR model. For each combination of parameters $t_1$, $t_2$, and $\mu$ we report the number of edges, $L$, minimum degree, $k_{\min}$, average degree, $\avg{k}$, maximum degree, $k_{\max}$, degeneracy, $D$, number of communities, $\NC$, and modularity, $Q$, for communities extracted using either the Louvain (\Louvain{}) or stochastic block model (\SBM{}) method. All networks have $N = 10000$ nodes.}
\label{tab:lfr_summary}
\end{table}
%

%
%
%
\begin{figure}[H]
\centering
\includegraphics[width=\linewidth]{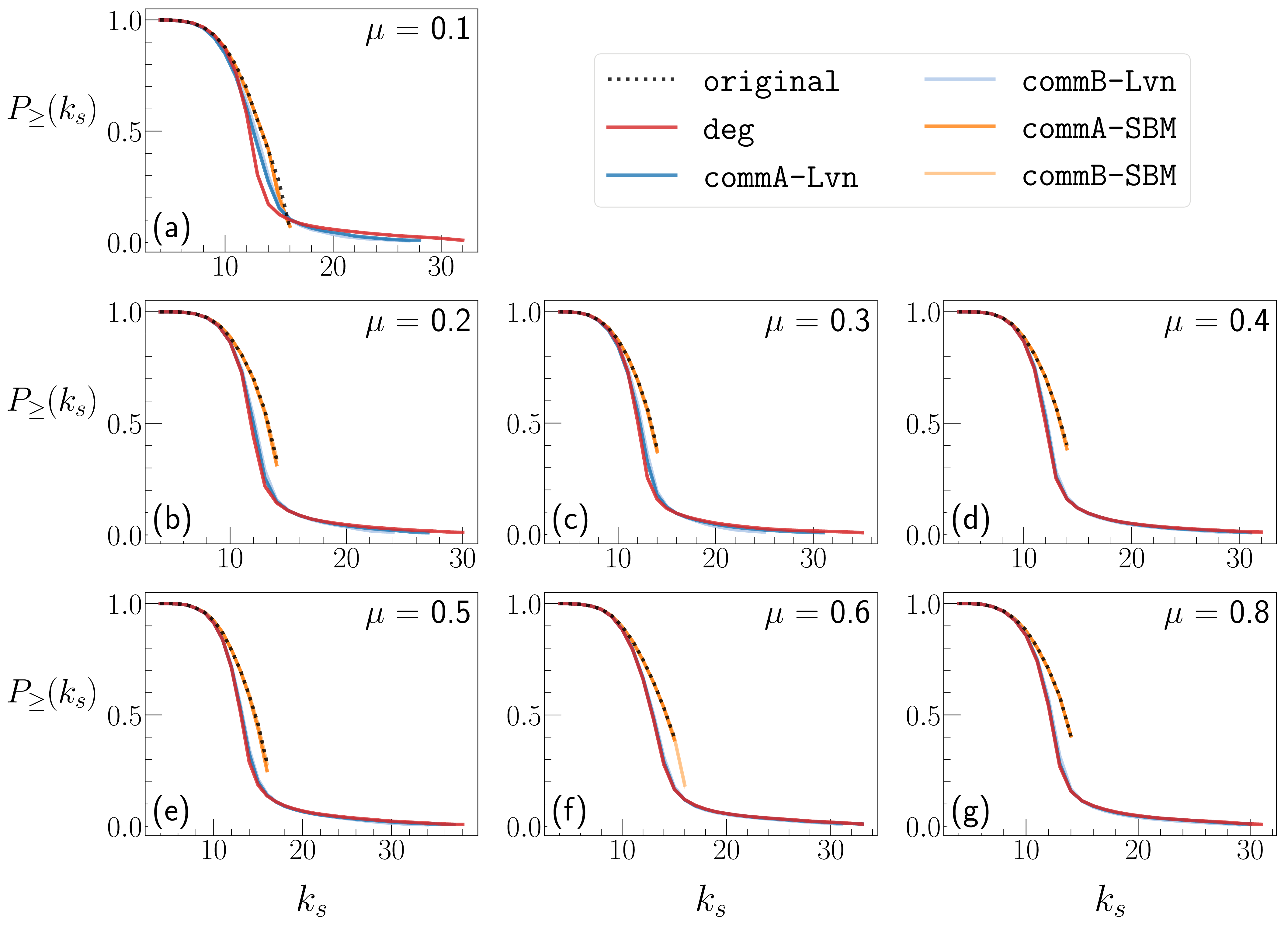}
\caption{Survival function of the probability distribution, $P_{\geq}(\ks)$, of the \kshell{} index, $\ks$, for the LFR networks generated using parameter batch 1 (\ie{} with $t_1 =2.2$ and $t_2 = 1.5$; see Supplementary Table~\ref{tab:lfr_summary}). The dotted lines correspond to the original network. The solid lines correspond to shuffled networks. Each panel corresponds to a value of $\mu$. Shuffled results are averages over $10$ realisations. The shaded area corresponds to the standard deviation. All networks have $N=10000$ nodes.}
\label{fig:cumR_LFR_G1}
\end{figure}
%

%
%
%
\begin{figure}[H]
\centering
\includegraphics[width=\linewidth]{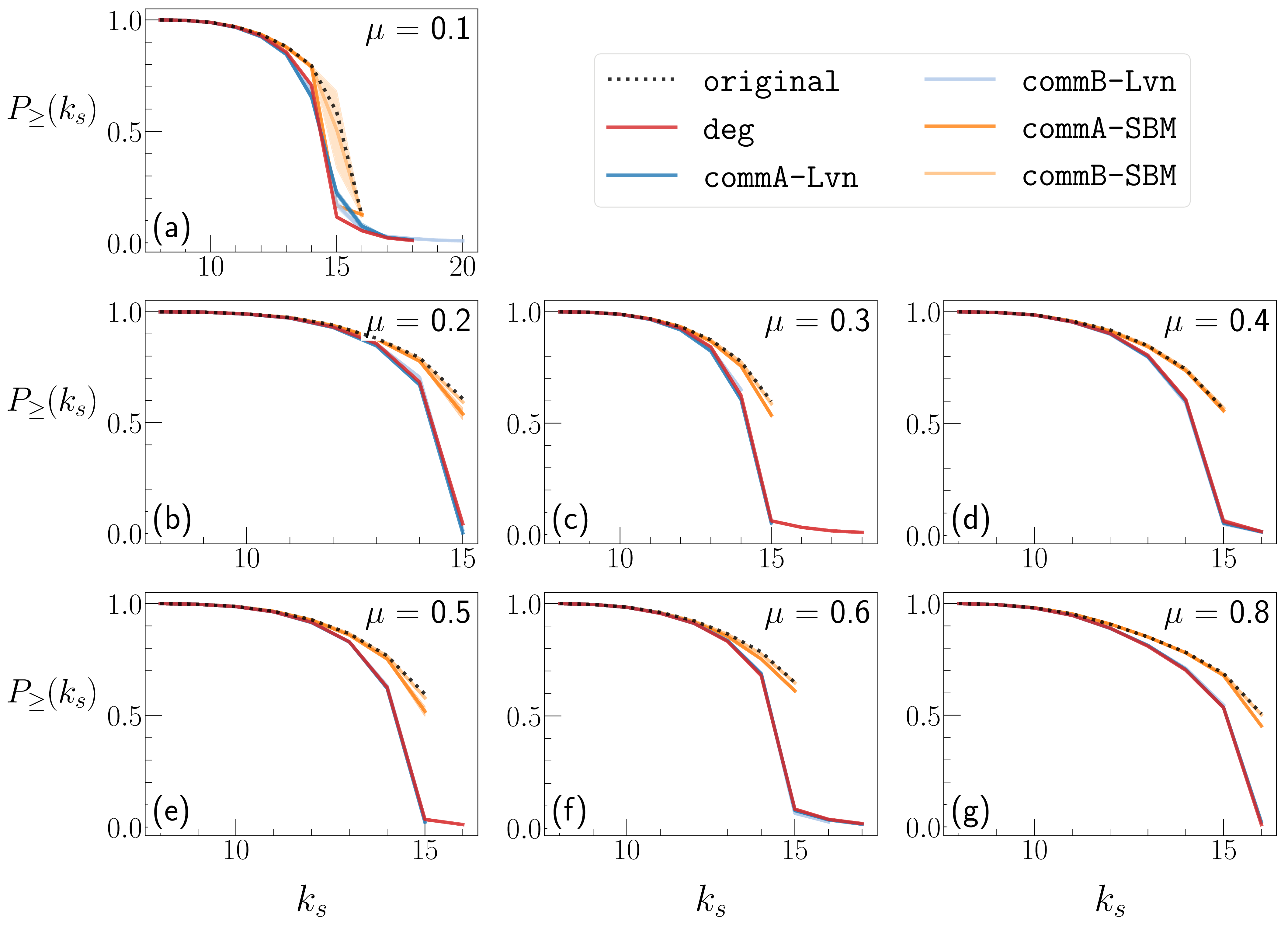}
\caption{Survival function of the probability distribution, $P_{\geq}(\ks)$, of the \kshell{} index, $\ks$, for the LFR networks generated using parameter batch 2 (\ie{} with $t_1 = 2.6$ and $t_2 = 2.0$; see Supplementary Table~\ref{tab:lfr_summary}). See the caption or Supplementary Fig.~\ref{fig:cumR_LFR_G1} for notations and legends.}
\label{fig:cumR_LFR_G2}
\end{figure}
%

%
%
%
\begin{figure}[H]
\centering
\includegraphics[width=\linewidth]{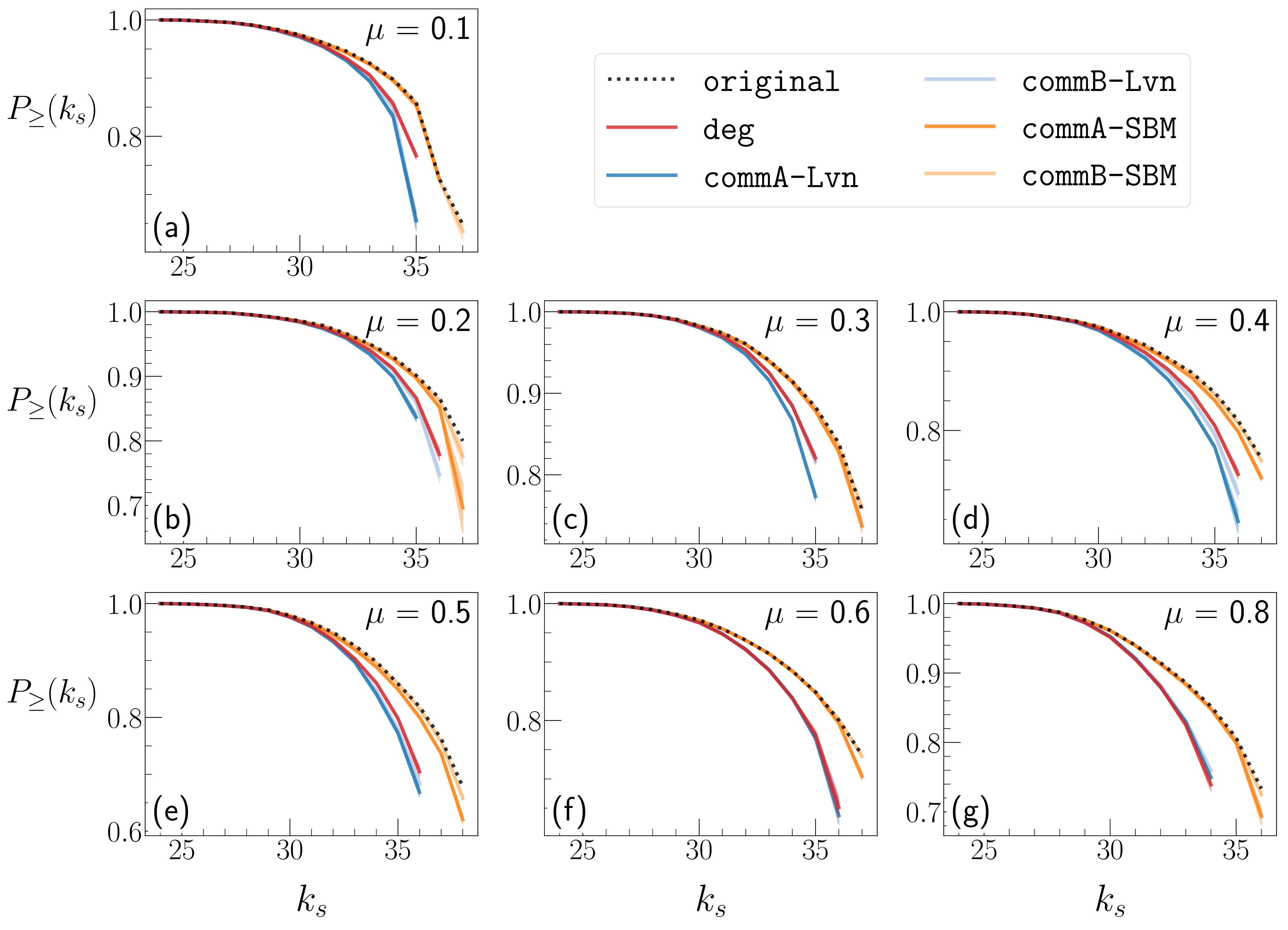}
\caption{Survival function of the probability distribution, $P_{\geq}(\ks)$, of the \kshell{} index, $\ks$, for the LFR networks generated using parameter batch 3 (\ie{} with $t_1 = 2.9$ and $t_2 = 1.5$; see Supplementary Table~\ref{tab:lfr_summary}). See the caption or Supplementary Fig.~\ref{fig:cumR_LFR_G1} for notations and legends.}
\label{fig:cumR_LFR_G3}
\end{figure}
%

%
%
%
\begin{figure}[H]
\centering
\includegraphics[width=\linewidth]{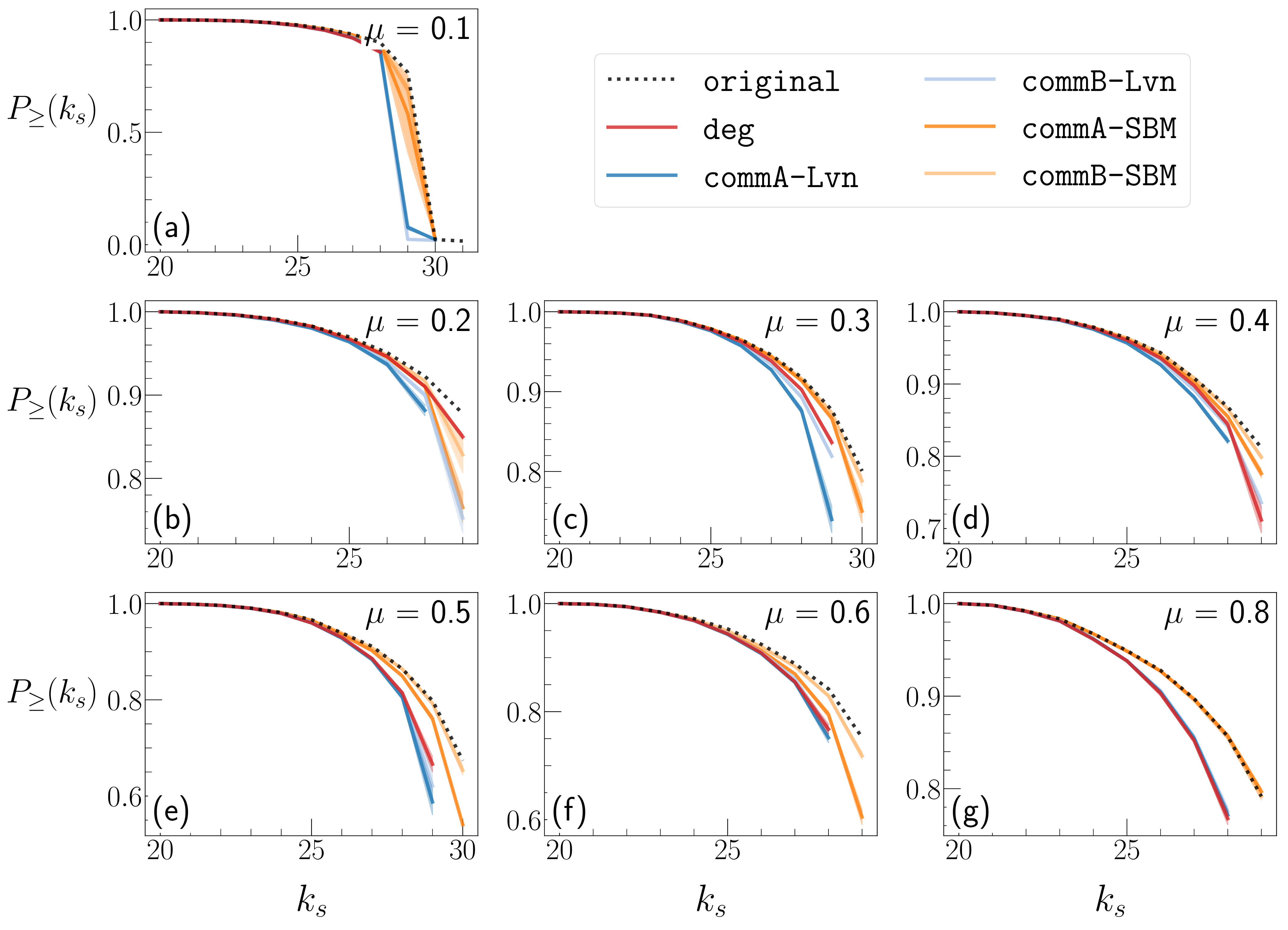}
\caption{Survival function of the probability distribution, $P_{\geq}(\ks)$, of the \kshell{} index, $\ks$, for the LFR networks generated using parameter batch 4 (\ie{} with $t_1 = 3.0$ and $t_2 = 2.0$; see Supplementary Table~\ref{tab:lfr_summary}). See the caption or Supplementary Fig.~\ref{fig:cumR_LFR_G1} for notations and legends.}
\label{fig:cumR_LFR_G4}
\end{figure}
%

%
%

\newpage

\section{Relationship between community structure and \kcore{} decomposition}
\label{sec:rel_comms_kcore}

\begin{figure}[H]
\centering
\includegraphics[width=0.85\linewidth]{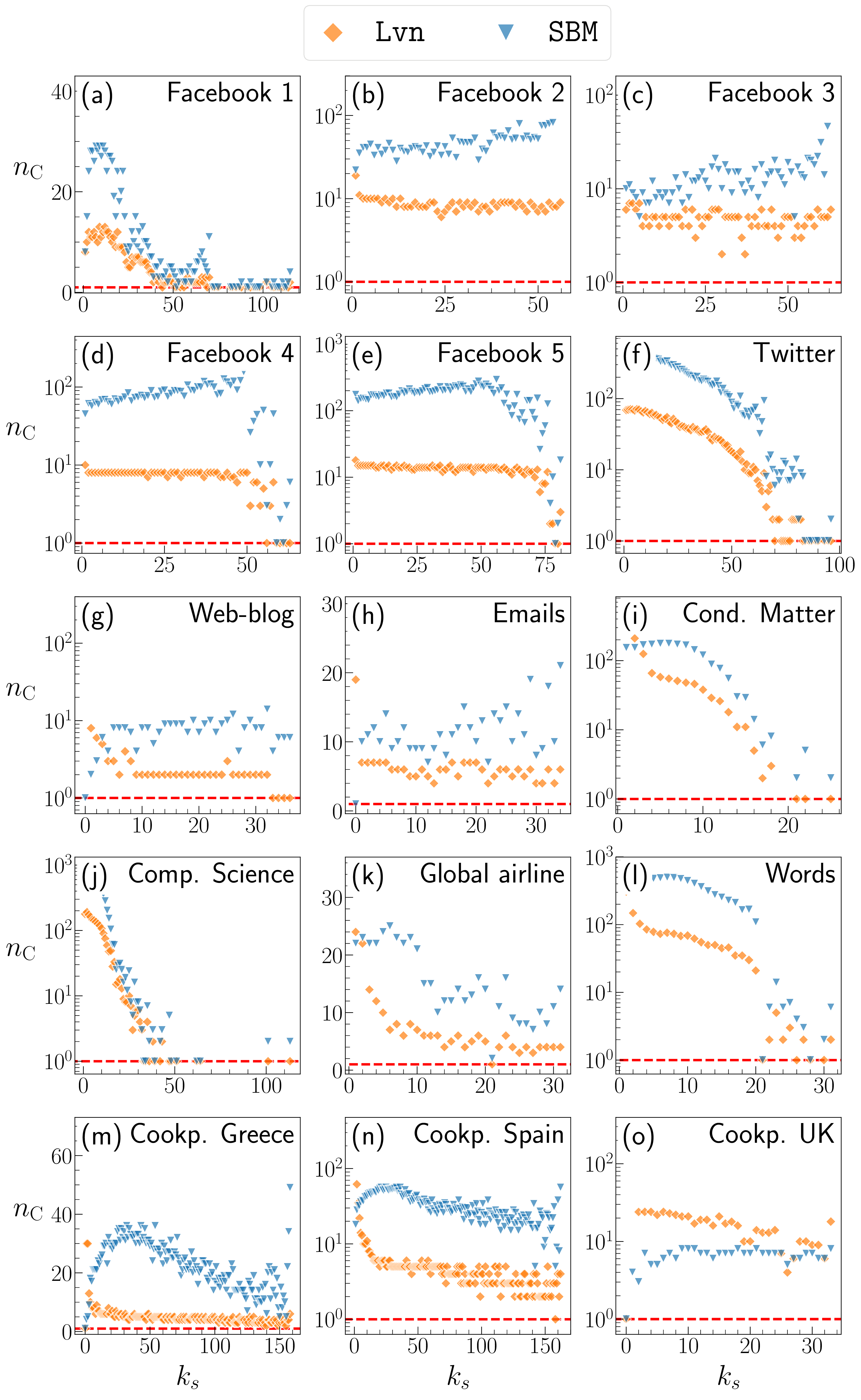}
\caption{Number of communities, $\nc(\ks)$, to which the nodes having \kshell{} index $\ks$ belong. The horizontal line is a guide to the eyes representing $\nc(\ks) = 1$. We identified the community structure using either \Louvain{} or \SBM{}. Each panel accounts for a different data set.}
\label{fig:nrcomms_vs_kshell_alldata}
\end{figure}

In this section, we examine the number of communities to which the nodes in each \kshell{} belong, with the aim of examining whether or not those nodes are concentrated into one or a small number of communities, particularly for nodes in innermost \kshell{s}. This analysis is similar to Supplementary Fig.~\ref{fig:mean_size_com}, whereas in that case we focused on the averaged community size. Supplementary Figure~\ref{fig:nrcomms_vs_kshell_alldata} shows the number of distinct communities to which the nodes with a given $\ks$ value belong, denoted by $\nc(\ks)$, for all the data sets. In agreement with Fig.~\ref{M-fig:nrcomms_vs_kshell}, some data sets show a strong concentration of the innermost \kshell{s} (\ie{} nodes with large $\ks$ values) into one or a few communities.

Next, we ask whether or not the number of communities across which each \kshell{} is distributed is a byproduct of random interactions. To answer this question, first, for each network, we extract communities using either \Louvain{} or \SBM{}. Second, we compute $\nc(\ks)$ for each $\ks$. Third, we compute the same quantity for the case in which we permute the association between the \kshell{} index of each node, $\ks(i)$, and the community membership of the node, $g(i)$, uniformly at random; in fact, it is sufficient to randomly permute either $\{ \ks(1), \ldots, \ks(N)\}$ or $\{g(1), \ldots, g(N)\}$, not both. Fourth, we calculate the number of communities to which the set of nodes with a given $\ks$ value belong after the permutation, which is denoted by $\nc^S(\ks)$. Fifth, using an approach similar to the calculation of the rich-club coefficient \cite{colizza-nphys-2006}, we compute
\begin{equation}
\label{eq:nc_ratio}
\varphi(\ks) = \dfrac{\nc^S(\ks)}{\nc(\ks)}
\end{equation}
for each $\ks$. A value of $\varphi(\ks)$ larger (smaller) than $1$ indicates that the number of communities to which the nodes having the $\ks$ value belong is smaller (larger) than in the case of the randomised association between the nodes and communities. Therefore, $\varphi(\ks)$ larger than $1$ implies that the nodes with the given \kshell{} index, $\ks$, are concentrated into a relatively small number of communities as compared to randomised counterparts.

In Supplementary Fig.~\ref{fig:ncomm_kshell_richclub_purecomb} we plot $\varphi(\ks)$ against $\ks$ for all the data sets. We observe that, with the exception of the Spanish and British Cookpad's networks, $\varphi(\ks)$ tends to be larger than $1$. This result implies that, on average, nodes of a given \kshell{} tend to belong to less communities than the randomised case.
We stress that the permutation of either the \kshell{} index or the community membership sequences may return networks whose \kshell{} and community structure are not physically plausible. For instance, if a node $i$ receives a \kshell{} index value of $\alpha$ upon randomisation and $\alpha$ is larger than $k_i$ (\ie{} degree of node $i$), then the node cannot belong to the corresponding \kshell{}.

%
%
%

%
%
%
\begin{figure}[H]
\centering
\includegraphics[width=.85\linewidth]{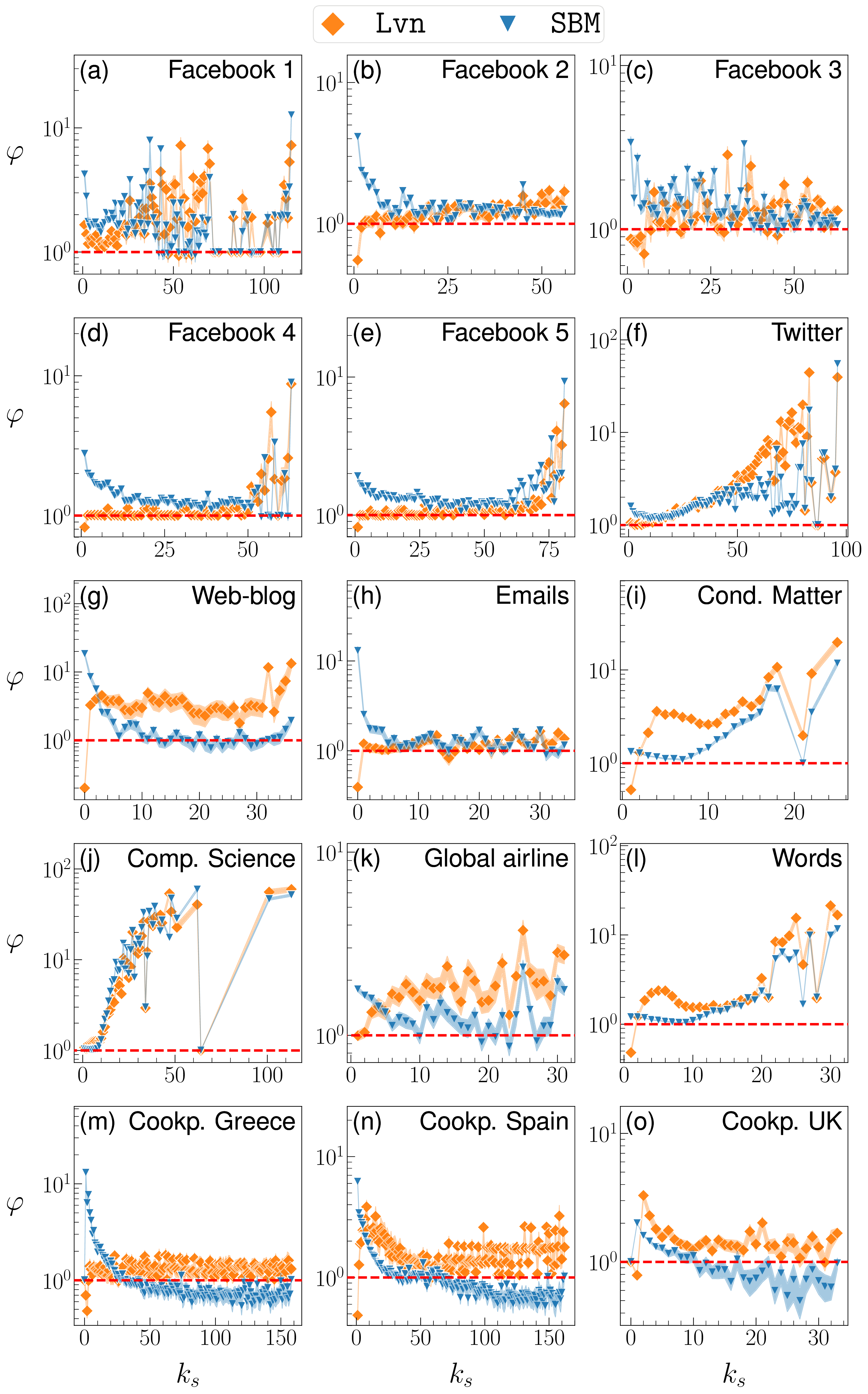}
\caption{Ratio, $\varphi(\ks)$, (see \eqref{eq:nc_ratio}) plotted against the \kshell{} index, $\ks$, for all the data sets. We identified the community structure using either \Louvain{} or \SBM{}. Each panel accounts for a different data set. Results are averaged over one hundred runs of randomisation between the association between the node's \kshell{} index and community label. The horizontal dashed lines represent $\varphi(\ks)=1$.}
\label{fig:ncomm_kshell_richclub_purecomb}
\end{figure}
%

%
%
\bibliographystyle{naturemag-doi}
\bibliography{biblio}

%
%

\makeatletter

\makeatother